\documentclass{cta-author}
\usepackage[linktocpage,colorlinks,linkcolor=blue,anchorcolor=blue,citecolor=blue,urlcolor=magenta,hypertexnames=false]{hyperref}

\usepackage{epstopdf}
\usepackage{graphicx}
\usepackage{float}
\usepackage{stfloats}
\usepackage{textcomp}
\usepackage{ifthen}
\usepackage{setspace}
\usepackage{multirow}
\usepackage{paralist}
\usepackage{algorithm}
\usepackage{algpseudocode}
\usepackage{algorithmicx}
\usepackage{array}
\usepackage{bm}
\usepackage{amsmath,amssymb,amsfonts,pifont}
\usepackage{graphicx}
\usepackage{textcomp}
\usepackage{subfig}
\usepackage[shortlabels,inline]{enumitem}
\usepackage{xcolor}

\usepackage[normalem]{ulem}
\useunder{\uline}{\ul}{}

\usepackage{xcolor,etoolbox}
\let\mybibitem\bibitem
\renewcommand{\bibitem}[1]{%
\ifstrequal{#1}{parizad2020security}{\color{black}\mybibitem{#1}}
{\ifstrequal{#1}{baghaee20161reliability}{\color{black}\mybibitem{#1}}
{\ifstrequal{#1}{baghaee2016power}{\color{black}\mybibitem{#1}}
{\ifstrequal{#1}{baghaee2017fuzzy}{\color{black}\mybibitem{#1}}
{\ifstrequal{#1}{belderbos2020facilitating}{\color{black}\mybibitem{#1}}
{\ifstrequal{#1}{mirzaei2020novel}{\color{black}\mybibitem{#1}}
{\ifstrequal{#1}{roald2020uncertainty}{\color{black}\mybibitem{#1}}
{\ifstrequal{#1}{conejo2020operations}{\color{black}\mybibitem{#1}}
{\ifstrequal{#1}{ameli2020coordinated}{\color{black}\mybibitem{#1}}
{\ifstrequal{#1}{dvorkin2020chance}{\color{black}\mybibitem{#1}}
{\color{black}\mybibitem{#1}} %
}}}}}}}}}}

{}
{}
{}
\newtheorem{proposition}{Proposition}{}
{}

\newenvironment{ldescription}[1]
  {\begin{list}{}%
   {\renewcommand\makelabel[1]{##1\hfill}%
   \settowidth\labelwidth{\makelabel{#1}}%
   \setlength\leftmargin{\labelwidth}
   \addtolength\leftmargin{\labelsep}}}
  {\end{list}}

\begin{document}
\supertitle{Submission Template for IET Research Journal Papers}

\title{Co-optimisation and Settlement of Power-Gas Coupled System in Day-ahead Market under Multiple Uncertainties}

\author{\au{Xiaodong~Zheng$^{1}$}, \au{Yan~Xu$^{2}$}, \au{Zhengmao~Li$^{2}$}, \au{Haoyong~Chen$^{1\corr}$}}

\address{\add{1}{School of Electric Power Engineering, South China University of Technology, Guangzhou 510641, China}
\add{2}{School of Electrical and Electronic Engineering, Nanyang Technological University, Singapore}
\email{eehychen@scut.edu.cn}}

\begin{abstract}
The interdependency of power systems and natural gas systems is being reinforced by the emerging power-to-gas facilities (PtGs), and the existing gas-fired generators. To jointly improve the efficiency and security under diverse uncertainties from renewable energy resources and load demands, it is essential to co-optimise these two energy systems for day-ahead market clearance. In this paper, a data-driven integrated electricity-gas system stochastic co-optimisation model is proposed. The model is accurately approximated by sequential mixed integer second-order cone programming, which can then be solved in parallel and decentralised manners by leveraging generalised Benders decomposition. \textcolor{black}{Since the price formation and settlement issues have rarely been investigated for integrated electricity-gas systems in an uncertainty setting}, a novel concept of expected locational marginal value is proposed to credit the flexibility of PtGs that helps hedging uncertainties. By comparing with a deterministic model and a distributionally robust model, the advantage of the proposed stochastic model and the efficiency of the proposed solution method are validated. Detailed results of pricing and settlement for PtGs are presented, showing that the expected locational marginal value can fairly credit the contribution of PtGs and reflect the system deficiency of capturing uncertainties.
\end{abstract}

\maketitle

\section{Introduction}\label{sec:introduction}
\subsection{\color{black}Motivation}
\textcolor{black}{Power-to-gas (PtG) is quite effective in storing large quantity of excess renewable electricity compared with conventional power-to-power energy storage technologies~\cite{simonis2017sizing}.}
{G}{iven} the high energy density of methane and the great potential of natural gas network as storages~\cite{blanco2018review}, PtG has been considered a promising technique in sustainable energy systems~\cite{bailera2017power, simonis2017sizing}.
Besides, natural gas-fired units (GfUs), {\color{black}despite being} traditional facilities, contribute an increasingly large share of the electricity generation~\cite{EIA_genshare, guo2018market}. The development of PtGs and the growth of GfUs tightly couple the electric power system with the natural gas system~\cite{chertkov2015cascading}.

{\color{black}The electric power system and the natural gas system are conventionally operated as individual systems without sufficient coordinations, as they belong to different energy sectors.} However, the intensified coupling has resulted in an integrated electricity-gas system (IEGS), for which coordinated operation become inevitable. Moreover, the liberalization of {\color{black}both the electricity market and the natural gas market~\cite{guo2018market, chinadaily2019pipeline, massrur2019hourly}}, together with the interactive safety and reliability {\color{black}requirements} of IEGS~\cite{correa2014integrated, chertkov2015cascading, belderbos2020facilitating}, are appealing for a security-constrained co-optimisation regime and corresponding settlement methods.

The challenges of co-optimizing IEGS in day-ahead markets include: \emph{i}) the uncertainties {\color{black}from} both renewable generations and electricity/gas demands, \emph{i}) the non-convexity of the natural gas flow {\color{black}model}, and \emph{iii}) the requirement of decentralised decision making. {\color{black}Therefore, it is necessary to develop a model that simultaneously addresses the above-mentioned issues with desired accuracy and reliability. Also, efficient solution algorithm should be developed.}

\textcolor{black}{Another practical challenge is the pricing issue or the settlement of these two energy sectors. Settlement of IEGS is a rather new topic, especially when the uncertainties of renewable generations and load demands are accounted for. Under an uncertainty environment, the traditional price formation mechanism in day-ahead markets must be systematically reevaluated and improved, because the original pricing regime may not be equitable and incentive enough for market participants who provide flexibilities and reserves.}

\subsection{\color{black}Literature Review}
\subsubsection{\color{black}Problem Modeling and Solution Algorithm}
The stochastic day-ahead scheduling problem of IEGS is investigated by~\cite{alabdulwahab2015coordination}, in which the natural gas flow problem is solved independently by Newton-Raphson substitution method to generate natural gas usage cuts. However, PtGs and the line-pack effect are ignored.
Ref.~\cite{bai2016interval} proposes an interval-optimisation-based model for IEGS to handle the wind power uncertainty, which is then solved directly by a mixed integer nonlinear programming (MINLP) solver.
A robust unit commitment (UC) model is developed in~\cite{he2017robust} to deal with the uncertainty of transmission line outage. Again, the demand uncertainty in the gas system is not considered, and both the line-pack effect and the gas compressor model are omitted in order for problem tractability.
Ref.~\cite{wang2018risk} deals with the optimal gas-power flow problem without considering the on/off statues of generators. Only the wind power uncertainty on the power system side is considered therein, which is addressed by distributionally robust optimisation.
{\color{black}Ref.~\cite{mirzaei2020novel} proposes a hybrid scenario and information gap based optimisation model for the day-ahead co-optimisation of IEGS under multiple uncertainties, and the MINLP is solved with a commercial solver.}
{\color{black}Ref.~\cite{roald2020uncertainty} proposes an uncertainty management framework for IEGS, which leveraging chance-constrained optimisation and robust optimisation. The transient gas pipeline flows are accurately modeled in~\cite{roald2020uncertainty}.}

To address the non-convexity of the problem and enable decentralised solutions, apart from the linearisation via Taylor series expansion~\cite{he2018co} and the second-order cone reformulation used in~\cite{wang2018risk, he2017robust}, Ref.~\cite{chen2019unit} proposes using mixed integer second-order cone constraints to enhance the approximation of the non-convex gas flow equation.
{\color{black}More recently, Ref.~\cite{ameli2020coordinated} proposes an outer approximation with equality relaxation method to cope with the non-convexity issue.}
In~\cite{zhao2018shadow}, the shadow price is utilised to coordinately optimize IEGS in day-ahead markets.
In the robust IEGS model of~\cite{he2016robust}, the non-convex natural gas problem is reformulated as a mixed integer linear programming (MILP), and the non-convex sub-problem of the robust optimisation model is solved distributedly via the alternating direction method of multipliers (ADMM) with heuristics.
In a subsequent work~\cite{wu2019distributionally}, the authors introduce price-sensitive demand-responses, and the uncertainty is handled by distributionally robust optimisation based on the linearised natural gas model. 

\subsubsection{\color{black}Pricing and Settlement}
Regarding the pricing and settlement issues, the authors in~\cite{chen2017clearing} propose a method for pricing the gas capacity reserved to GfUs. However, the non-convex gas transmission constraints are approximated by some linear cuts, and constraints in stochastic scenarios are discarded.
The strategic offering and equilibrium problem of coordinated IEGS markets is investigated in~\cite{wang2017strategic}, whereas the line-pack effect and gas nodal pressure variables are omitted for problem tractability.
A scenario-based model is proposed in~\cite{li2018optimal} to determine the optimal {\color{black}operation} strategy of GfUs and PtGs in energy and regulation markets. Further, the Shapley value is employed to allocate the payoff among these facilities.

The concept of cost of uncertainty is developed in~\cite{zhang2014network} to characterize the impact of uncertainty on the dispatch cost, but the value of flexible resources is not evaluated.
In~\cite{ye2016uncertainty}, the authors make use of the derivative of a robust UC model to construct the uncertainty marginal price, which quantifies the value of reserve and the cost of uncertainty in the day-ahead market.
A recent work in~\cite{fang2019introducing} deals with the problem of pricing transmission overload and generation violation caused by random renewable generations and demands. Therein, a distributionally robust chance-constrained optimal power flow model is developed, {\color{black}which renders uncertainty-contained locational marginal prices that determine} how the revenue should be distributed to conventional generators.
{\color{black}More recently, Ref.~\cite{dvorkin2020chance} proposes a chance-constrained stochastic pricing method for linear electricity markets, in which the price is formed by a scenario-independent mathematical programming reduced from the chance-constrained model.}

\subsection{\color{black}Contribution and Paper Organization}
In this paper, a day-ahead co-optimisation problem of IEGS is investigated, considering the uncertainties of both renewable generations and electricity/gas demands. Moreover, the price formation and settlement issue is studied with a focus on PtGs, and the economic efficiency of PtGs is also analysed. The proposed co-optimisation method and settlement regime are validated by thorough numerical results and comparisons with a deterministic model and a distributionally robust model.

{\color{black}The detailed technical} contributions of this paper include:
\begin{enumerate}[1)]
  \item {\color{black}A stochastic day-ahead market model} is developed for the integrated electricity-gas system, which precisely accounts for the natural gas flow constraints, line-pack effect, PtGs, as well as correlated uncertainties. {\color{black}The stochastic model makes use of data-driven scenarios so that the natures of multiple uncertainties could be better retained.}
  \item The stochastic model is approximated by sequential mixed integer second-order cone programming (MISOCP), which is shown to be highly precise. Based on generalised Benders decomposition, the convex sub-problems are further decoupled and solved by the electric power system operator and the natural gas system operator decentrally. The stochastic model and the ensemble solution method are shown to have advantages over state-of-the-arts in terms of dealing with the uncertainty, the non-convexity, and the decentralised decision making issues.
  \item {\color{black}A novel concept of expected locational marginal value (E-LMV) is proposed for price formation in the electricity-gas market, which has advantages in crediting PtGs equally and ensuring cost recovery of such flexibility providers in a power-gas coupled market with production and demand uncertainties}. Moreover, the revenue adequacy condition of the day-ahead natural gas market is analysed for the first time.
\end{enumerate}


\textcolor{black}{The remainder of this paper is organised as follows: Section~\ref{sec:GS} and Section~\ref{sec:SCUC} introduce the natural gas system model, and the electric power system model, respectively. Section~\ref{sec:IEGS} establishes the stochastic model for the power-gas coupled system, and introduces the novel pricing method. Section~\ref{sec:alg} presents the solution algorithms. Numerical experiments and detailed results are reported in Section~\ref{sec:case}. Section~\ref{sec:conclusion} concludes with discussions.}

\section{Natural Gas System Model}\label{sec:GS}
This section {presents} a \textcolor{black}{dynamic/multi-period} optimal flow model of the natural gas system. Typical components are modeled including gas compressors, gas storages, GfUs and PtGs. The gas traveling velocity and compressibility are accounted for~\cite{correa2014integrated}, as gas travels much slower than electricity and it can be stored in pipelines. \textcolor{black}{Further, we assume by convention that the state variables of the natural gas system are stable within each 1-hour scheduling time slot~\cite{he2018coordination}.}
\subsection{{Notation for Natural Gas system }}
\begin{ldescription}{$xxxxxxxxx$}
    \item [$\mathcal{N},\mathcal{T}$] Set of nodes in the natural gas system and set of scheduling time periods.
    \item [$\mathcal{G}^{\rm{Src}}_{n},\mathcal{G}^{\rm{Str}}_{n}$] Sets of natural gas suppliers and gas storages, connected at node $n$.
    \item [$\mathcal{G}^{\rm{Cmp}}_{n},\mathcal{G}^{\rm{Pipe}}_{n}$] Sets of active pipelines (with gas compressors) and passive pipelines (without gas compressors) connected with node $n$; node $n$ is the outlet of pipeline $(m,n)$, or the inlet of pipeline $(n,m)$.
    \item [$\mathcal{G}^{\rm{GfU}}_{n},\mathcal{G}^{\rm{PtG}}_{n}$] Sets of GfUs and PtGs connected at node $n$.
    \item [$\mathcal{G}^{\rm{Load}}_{n}$] Set of non-generation-related natural gas demands connected at node $n$.
    \item [$\mathcal{G}^{\rm{Pipe}}$] Set of all directed passive pipelines that have positive gas flow.
    \item [$P^{\rm{Src}}_w,P^{\rm{Str}}_s$] Price of natural gas from supplier $w$ and price of gas for gas storage station $s$ [\$/Mscm]\footnote{Mscm means million standard cubic meters of gas.}.
    \item [{$\underline{F}^{\rm{Src}}_w,\overline{F}^{\rm{Src}}_w$}] Lower and upper flow limits of natural gas from supplier $w$ [Mscm/h].
    \item [$\underline{G}_w,\overline{G}_w$] Limits of daily quantity from supplier $w$ according to gas-delivery contracts [Mscm].
    \item [{$\underline{F}^{\rm{Str}}_s,\overline{F}^{\rm{Str}}_s$}] Outflow and inflow limits of storage station $s$ [Mscm/h].
    \item [$\underline{S}_s,\overline{S}_s$] Capacity limits of storage station $s$ [Mscm].
    \item [$S^{\rm{Str}}_{s,0}$] Initial gas volume in storage station $s$ [Mscm].
    \item [{$\underline{\Pi}_{n},\overline{\Pi}_{n}$}] Gas pressure limits of node $n$ [bar].
    \item [$\underline{C}/\overline{C}_{(m,n)}$] Compression ratio limits of compressor at pipeline $(m,n)$.
    \item [$\beta(m,n)$] Efficiency factor of compressor at pipeline $(m,n)$.
    \item [$\Delta$] Time slot in the scheduling models [1 h].
    \item [$K^{\rm{gf}}_{(m,n)}$] Natural gas flow constant [Mscm/(h$\cdot$bar)].
    \item [$K^{\rm{lp}}_{(m,n)}$] Line-pack constant [Mscm/bar].
    \item [$E_{(m,n),0}$] Initial line pack of pipeline $(m,n)$ [Mscm].
    {\item [${\rm{sgn}}(\cdot)$] Sign function that returns -1 for negative input, 0 for zero, and 1 for positive input.}
    \item [$f^{\rm{Src}}_{w,t}$] Natural gas supplied by supplier $w$ at time $t$ [Mscm/h].
    \item [$f^{\rm{Str}}_{s,t}$] Natural gas flows into gas storage $s$ at time $t$ [Mscm/h]; negative if the real gas flows out.
    \item [$\pi_{n,t}$] Gas pressure of node $n$ at time $t$ [bar].
    \item [$\delta^{\rm{Cmp}}_{(m,n),t}$] Natural gas consumed by compressor at pipeline $(m,n)$ [Mscm/h].
    \item [$f^{{\rm{Cmp}}/{\rm{Pipe}}}_{(m,n),t}$] Gas flows through active/passive pipeline $(m,n)$ at time $t$ [Mscm/h]; negative if the real gas flow reaches node $m$ from pipeline $(m,n)$.
    \item [$\bar{f}^{\rm{Pipe}}_{(m,n),t}$] Average gas flow in pipeline $(m,n)$ at time $t$ [Mscm/h].
    \item [{$e_{(m,n),t}$}] Line pack of pipeline $(m,n)$ at time $t$ [Mscm].
    \item [$f^{\rm{GfU}}_{g,t}$] Gas consumption of GfU $g$ at time $t$ [Mscm/h].
    \item [$f^{\rm{PtG}}_{v,t}$] Gas production of PtG $v$ at time $t$ [Mscm/h].
    \item [{$F^{\rm{Load}}_{s,t}$}] Non-generation-related natural gas demand at node $n$ at time $t$ [Mscm/h].
\end{ldescription}\vspace{-2mm}

\subsection{Model Formulation}
The GS model is formulated as:
\begin{subequations} \label{eqn:GS_full}
    \begin{align}
        \nonumber & \min\, \sum\limits_{w\in\mathcal{G}^{\rm{Src}}_{n}} \sum\limits_{t\in\mathcal{T}} {P^{\rm{Src}}_w f^{\rm{Src}}_{w,t} \Delta} - \sum\limits_{s\in\mathcal{G}^{\rm{Str}}_{n}} \sum\limits_{t\in\mathcal{T}} {P^{\rm{Str}}_s f^{\rm{Str}}_{s,t} \Delta} \\
        \nonumber & \operatorname{s.t.}~~ \forall n\in\mathcal{N} \\
        & {\underline{F}^{\rm{Src}}_w} \le f^{\rm{Src}}_{w,t} \le {\overline{f}^{\rm{Src}}_w} ~~\forall w\in\mathcal{G}^{\rm{Src}}_{n},t\in\mathcal{T} \label{eqn:GS_src_flow_limit}\\
        & \underline{G}_w \le \sum_{t\in\mathcal{T}}f^{\rm{Src}}_{w,t} \le \overline{G}_w ~~\forall w\in\mathcal{G}^{\rm{Src}}_{n} \label{eqn:GS_src_quantity_limit}\\
        & {\underline{F}^{\rm{Str}}_s} \le f^{\rm{Str}}_{s,t} \le {\overline{F}^{\rm{Str}}_s} ~~\forall s\in\mathcal{G}^{\rm{Str}}_{n},t\in\mathcal{T}\label{eqn:GS_str_flow_limit}\\
        & \underline{S}_s \le S^{\rm{Str}}_{s,0} + \sum^{t}_{\tau=1}f^{\rm{Str}}_{s,\tau} \le \overline{S}_s ~~\forall s\in\mathcal{G}^{\rm{Str}}_{n},t\in\mathcal{T} \label{eqn:GS_str_capicity_limit}\\
        & {\underline{\Pi}_{n}} \le \pi_{n,t} \le {\overline{\Pi}_{n}} ~~\forall t\in\mathcal{T} \label{eqn:GS_pi_limit}\\
        & \underline{C}_{(m,n)}\pi_m \le \pi_n \le \overline{C}_{(m,n)}\pi_m ~\forall (m,n)\in\mathcal{G}^{\rm{Cmp}}_{n},t\in\mathcal{T} \label{eqn:GS_cmp_rate}\\
        & \delta^{\rm{Cmp}}_{(m,n),t} = \beta_{(m,n)}|f^{\rm{Cmp}}_{(m,n),t}| ~~\forall (m,n)\in\mathcal{G}^{\rm{Cmp}}_{n},t\in\mathcal{T} \label{eqn:GS_cmp_consume}\\
        \nonumber & {e_{(m,n),t}} = \Delta\cdot \left(f^{\rm{Cmp}}_{(m,n),t}+f^{\rm{Cmp}}_{(n,m),t}-\delta^{\rm{Cmp}}_{(m,n),t}\right) \\
        & ~~~~~~~~~~~~~ + {e_{(m,n),t-1}} ~~\forall (m,n)\in\mathcal{G}^{\rm{Cmp}}_{n},{t}\in\mathcal{T} \label{eqn:GS_cmp_lp_f}\\
        \nonumber & \bar{f}^{\rm{Pipe}}_{(m,n),t} = {\rm{sgn}}(\pi_{m,t},\pi_{n,t})K^{\rm{gf}}_{(m,n)}\sqrt{|\pi^2_{m,t}-\pi^2_{n,t}|} \\
        & ~~~~~~~~~~~~~~~~~~~~~~~~~~~~~~~~~~~~\forall (m,n)\in\mathcal{G}^{\rm{Pipe}}_{n},t\in\mathcal{T} \label{eqn:GS_f_Pipe_bar_pi}\\
        & {\bar{f}}^{\rm{Pipe}}_{(m,n),t} = \left(f^{\rm{Pipe}}_{(m,n),t}-f^{\rm{Pipe}}_{(n,m),t}\right)/2 ~\forall (m,n)\in\mathcal{G}^{\rm{Pipe}}_{n},t\in\mathcal{T} \label{eqn:GS_f_Pipe_bar_f}\\
        & {e_{(m,n),t}} = K^{\rm{lp}}_{(m,n)}\left(\pi_{m,t}+\pi_{n,t}\right)/2 ~\forall (m,n)\in\mathcal{G}^{\rm{Pipe}}_{n},t\in\mathcal{T} \label{eqn:GS_lp_pi}\\
        \nonumber & {e_{(m,n),t}} = \Delta\cdot \left(f^{\rm{Pipe}}_{(m,n),t}+f^{\rm{Pipe}}_{(n,m),t}\right) \\
        & ~~~~~~~~~~~~~ + {e_{(m,n),t-1}}~~\forall (m,n)\in\mathcal{G}^{\rm{Pipe}}_{n},{t}\in\mathcal{T} \label{eqn:GS_lp_f}\\
        & {e_{(m,n),|\mathcal{T}|}} = E_{(m,n),0} ~~\forall (m,n)\in\mathcal{G}^{\rm{Cmp}}_{n}\cup\mathcal{G}^{\rm{Pipe}}_{n} \label{eqn:GS_lp_identity}\\
        \nonumber & f^{\rm{Src}}_{{w,t}|{w\in\mathcal{G}^{\rm{Src}}_n}} + f^{\rm{PtG}}_{v,t|{v\in\mathcal{G}^{\rm{PtG}}_n}} = f^{\rm{Str}}_{s,t|{s\in\mathcal{G}^{\rm{Str}}_n}} + f^{\rm{GfU}}_{g,t|{g\in\mathcal{G}^{\rm{GfU}}_n}} \\
        \nonumber &  ~~~~~~~~~~~~~~~~~~~+ {F^{\rm{Load}}_{d,t|{d\in\mathcal{G}^{\rm{Load}}_n}}} + f^{\rm{Cmp}}_{(n,m),t|{(n,m)\in\mathcal{G}^{\rm{Cmp}}_n}} \\
        & ~~~~~~~~~~~~~~~~~~~~~~~~~~~~+f^{\rm{Pipe}}_{(n,m),t|{(n,m)\in\mathcal{G}^{\rm{Pipe}}_n}}~\forall {t}\in\mathcal{T} \label{eqn:GS_flow_balance}.
    \end{align}
\end{subequations}
{The objective function accounts for the gas volume from suppliers and the net gas consumption of storages.}
Constraints~(\ref{eqn:GS_src_flow_limit}) and (\ref{eqn:GS_src_quantity_limit}) define flow limits and daily quantity limits of gas sources. Constraints~(\ref{eqn:GS_str_flow_limit}) and (\ref{eqn:GS_str_capicity_limit}) define flow limits and capacity limits of gas storages. Constraint~(\ref{eqn:GS_pi_limit}) restricts the gas pressure of each node to be within a safety range. For active pipelines, compression ratios are limited by constraint~(\ref{eqn:GS_cmp_rate}), while gas consumptions and line packs are defined respectively by constraints~(\ref{eqn:GS_cmp_consume}) and (\ref{eqn:GS_cmp_lp_f}). For passive pipelines, the general flow equation~(\ref{eqn:GS_f_Pipe_bar_pi}) expresses the relationship between the pressure gradient and the gas flow, which can be evaluated via~(\ref{eqn:GS_f_Pipe_bar_f}); Equality~(\ref{eqn:GS_lp_pi}) indicates that the line pack is proportional to the average pressure, and the line pack should also complies with the mass conservation~(\ref{eqn:GS_lp_f}). Constraint~(\ref{eqn:GS_lp_identity}) imposes a requirement on line-pack level in the last scheduling period. Constraint~(\ref{eqn:GS_flow_balance}) enforces gas balance at each node.

{Constraint~(\ref{eqn:GS_cmp_consume}) adopts a simplified gas consumption function for the compressor~\cite{he2018coordination, chen2019unit} instead of the original one, which is highly nonlinear on the gas flow through and the compression ratio~\cite{he2018coordination}.}
Constraints~(\ref{eqn:GS_f_Pipe_bar_pi}) and (\ref{eqn:GS_lp_pi}) can also be applied to active pipelines after such pipelines are separated into two segments from the location of compressors, but this is necessary only when the length of a pipeline is considerable.
Two key parameters of the natural gas flow model, i.e., $K^{\rm{gf}}_{(m,n)}$ and $K^{\rm{lp}}_{(m,n)}$, are calculated according to the equations derived {in the appendix of}~\cite{correa2014integrated}. To obtain $K^{\rm{lp}}_{(m,n)}$, the friction factor of pipeline is yielded from the Nikuradse equation first{, which is detailed in~\cite{de2000gas}. Parameters used to calculate $K^{\rm{gf}}_{(m,n)}$ and $K^{\rm{lp}}_{(m,n)}$ are available online~\cite{zheng_2019}.}

\section{Electric Power System Model}\label{sec:SCUC}
This section {presents a basic} security-constrained unit commitment (SCUC) model {for the electric power system}. The reserve requirements are omitted herein since stochastic programming is used in this paper. Nevertheless, constraints for the reserve are retained in a deterministic model, which is adopted as benchmark in case studies.
\subsection{{Notation for Electric Power System}}
\begin{ldescription}{$xxxxxxxx$}
    \item [$\mathcal{G},\mathcal{G}^{\rm{CfU}}$] Sets of all units and coal-fired units (CfUs).
    \item [$\mathcal{L}$] Set of transmission lines.
    \item [$\mathcal{E},\mathcal{E}_{\rm{ref}}$] Sets of buses and reference bus.
    \item [$NL_{g,t}$] No-load cost of generator $g$ at time $t$ [\$].
    \item [$SU_{g,t}$] Start-up cost of generator $g$ at time $t$ [\$].
    \item [$SD_{g,t}$] Shut-down cost of generator $g$ at time $t$ [\$].
    \item [$C_{g,t}$] Variable cost of generator $g$ at time $t$ [\$/MWh].
    \item [$MU_{g}$] Minimum-up time of unit $g$ [h].
    \item [$MD_{g}$] Minimum-down time of unit $g$ [h].
    \item [$R^{+/-}_{g}$] Ramp-up/ramp-down limit of unit $g$ [MW/h].
    \item [$X_{(m,n)}$] Reactance of transmission line $(m,n)$ [kV\textsuperscript{-2}$\Omega$].
    \item [$\overline{F}_{(m,n)}$] Rating of transmission line $(m,n)$ [MW].
    \item [$B_{m,n}$] Element on the $m$-row and the $n$-th column of the nodal susceptance matrix [kV\textsuperscript{2}S].
    \item [$\kappa_{g,n}^{\mathcal{G}},\kappa_{v,n}^{\rm{PtG}}$] 0-1 coefficient indicating whether unit $g$ or PtG $v$ is connected at bus $n$.
    \item [{$\underline{P}_{g},\overline{P}_{g}$}] Minimum and maximum production levels of unit $g$ [MW].
    \item [$x_{g,t}$] Binary variables indicating whether the unit is on.
    \item [$u_{g,t},v_{g,t}$] Binary variables indicating whether the unit is started up and shut down.
    \item [$p_{g,t}$] Production level of unit $g$ at time $t$ [MW].
    \item [$\theta_{n,t}$] Phase angel of bus $n$ at time $t$ [rad].
    \item [$p^{\rm{GfU}}_{g,t}$] Power output of GfU $g$ at time $t$ [MW].
    \item [$p^{\rm{PtG}}_{v,t}$] Power consumption of PtG $v$ at time $t$ [MW].
    \item [{$D_{n,t}$}] Load demand at bus $n$ at time $t$ [MW].
    \item [{$W_{n,t}$}] Output of renewable energy sources (RES) at bus $n$ at time $t$ [MW].
\end{ldescription}\vspace{-2mm}

\subsection{Model Formulation}
The SCUC model is formulated as:
\begin{subequations} \label{eqn:SCUC}
    \begin{align}
        \nonumber & \min\, \sum\limits_{g\in {\mathcal{G}}^{\rm{CfU}}}{\sum\limits_{t\in \mathsf{\mathcal{T}}}{{{x}_{g,t}}N{{L}_{g,t}}+{{u}_{g,t}}S{{U}_{g,t}}}} \\
        \nonumber & ~~~~~~~~~~~~~~~~~~~+{{v}_{g,t}}S{{D}_{g,t}}+{{C}_{g,t}}{{p}_{g,t}}\Delta \\
        \nonumber & \operatorname{s.t.} \\
        & {{x}_{g,t}}-{{x}_{g,t-1}}={{u}_{g,t}}-{{v}_{g,t}} ~~\forall g\in\mathcal{G}^{\rm{CfU}},{t,t-1}\in\mathcal{T} \label{eqn:SCUC_status} \\
        & \sum\limits_{\tau =\max \left\{ 1,t-M{{U}_{g}}+1 \right\}}^{t}{{{u}_{g,\tau }}}\le {{x}_{g,t}} ~~\forall g\in\mathcal{G}^{\rm{CfU}},t\in\mathcal{T} \label{eqn:SCUC_min_on} \\
        & \sum\limits_{\tau =\max \left\{ 1,t-M{{D}_{g}}+1 \right\}}^{t}{{{v}_{g,\tau }}}\le 1-{{x}_{g,t}} ~~\forall g\in\mathcal{G}^{\rm{CfU}},t\in\mathcal{T} \label{eqn:SCUC_min_down} \\
        & {\color{black}{\underline{P}_{g}}{x}_{g,t} \le {{p}_{g,t}} \le {\overline{P}_{g}}{x}_{g,t} ~~\forall g\in\mathcal{G},t\in\mathcal{T}} \label{eqn:SCUC_gen_limit} \\
        & R_{g}^{-}\le {{p}_{g,t}}-{{p}_{g,t-1}}\le R_{g}^{{+}} ~~\forall g\in {{\mathsf{\mathcal{G}}}},t,t-1\in \mathsf{\mathcal{T}} \label{eqn:SCUC_ramping_limit} \\
        & \left|(\theta_{n,t}-\theta_{m,t})/X_{(m,n)}\right| \le \overline{F}_{(m,n)} ~~\forall (m,n)\in\mathcal{L}, t\in\mathcal{T} \label{eqn:SCUC_flow_limit}\\
        & \theta_{n,t} = 0 ~~ n\in\mathcal{E}_{\rm{ref}} \label{eqn:SCUC_ref} \\
        \nonumber & \sum_{m\in\mathcal{E}}B_{n,m}\theta_{m,t} = p_{g,t|\kappa_{g,n}^{\mathcal{G}}=1} + {W_{n,t}} - {D_{n,t}} - p^{\rm{PtG}}_{v,t|\kappa^{\rm{PtG}}_{v,n}=1} \\
        & ~~~~~~~~~~~~~~~~~~~~~~~~~~~~~~~~~~~~~~~~~~~~~~\forall n\in\mathcal{E}, t\in\mathcal{T}. \label{eqn:SCUC_dc_equation}
    \end{align}
\end{subequations}
{The objective function accounts for the start-up and shut-down costs of CfUs, and the generation cost of CfUs.}
Constraints~(\ref{eqn:SCUC_status})-(\ref{eqn:SCUC_min_down}) include state transition equations of units and minimum up/down time limits of units. Constraints~(\ref{eqn:SCUC_gen_limit})-(\ref{eqn:SCUC_flow_limit}) are production limits of units, ramping limits of units, and power flow limits of transmission lines respectively. Equation~(\ref{eqn:SCUC_ref}) designates a reference bus, and {the dc power flow equation}~(\ref{eqn:SCUC_dc_equation}) enforces power balance at each bus.

In the electric power system model, the start-up and shut-down costs, as well as the on/off variables of GfUs are omitted.
{This is due to two facts. First, GfUs are quick-start units that can change their intra-day on/off statues, so it is inappropriate to fix their statues day-ahead. Second, the on/off statues can be ignored in the optimisation model without affecting the engineering behavior of GfUs (because the start-up time and minimum production level of GfUs are quite short/low), while incorporating binary variables into the convex dispatch problem will complicate the stochastic counterpart of this problem a lot (e.g., the dispatch problem becomes a mixed integer programming, to which many decomposition algorithms are no longer applicable).}
In practice, we can simply add a constant term to the objective function to account for the daily average start-up and shut-down costs of GfUs though.

{It is worth mentioning that in the implementation of the models, slack variables indicating load shedding and renewable generation curtailment are introduced to the gas/power balance equations, and the penalty costs are augmented to the objective functions accordingly.}

\section{Modeling for Integrated Electricity-Gas System with Multiple Uncertainties}\label{sec:IEGS}

\subsection{Integrated Electricity-Gas System}
It is assumed that the generators can be divided into two groups, i.e., CfU and GfU. Thus, we have $\mathcal{G}^{\rm{GfU}}= \mathcal{G} \backslash \mathcal{G}^{\rm{CfU}}$. Besides, we have $\mathcal{G}^{\rm{PtG}} = {\bigcup_{n\in\mathcal{N}}} \mathcal{G}^{\rm{PtG}}_{n}$ for PtG facilities. The natural gas system and the power system are coupled via the following equations:
\begin{equation} \label{eqn:GasPower_coupling} 
\left\{ \begin{aligned}
        & p^{\rm{GfU}}_{g,t}=\eta^{\rm{GfU}}_{g} f^{\rm{GfU}}_{g,t} H_{\rm{g}} ~~\forall g\in\mathcal{G}^{\rm{GfU}},t \in\mathcal{T} \\
        & f^{\rm{PtG}}_{v,t}=\eta^{\rm{PtG}}_{v} p^{\rm{PtG}}_{v,t} / H_{\rm{g}} ~~\forall v\in\mathcal{G}^{\rm{PtG}},t \in\mathcal{T},  \\
    \end{aligned} \right.
\end{equation}
where $\eta^{\rm{GfU}}_{g},\eta^{\rm{PtG}}_{v}$ are the efficiencies of GfU $g$ and PtG $v$ given by 0.43 and 0.58 respectively~\cite{wei2017power,wang2018risk}, and $H_{\rm{g}}$ is the heating rate of natural gas given by 1.08$\times$10\textsuperscript{4} MW/Mscm.

The coupling parameters are regarded as decision variables in IEGS, so it is necessary to add bounds for them, e.g.,
\begin{equation}\label{eqn:coupling_var_bound}
  0 \le p^{\rm{GfU}}_{g,t} \le {\overline{p}^{\rm{GfU}}_{g}},~ 0 \le p^{\rm{PtG}}_{v,t} \le {\overline{p}^{\rm{PtG}}_{v}}
\end{equation}
where ${\overline{p}^{\rm{GfU}}_{g}},{\overline{p}^{\rm{PtG}}_{v}}$ are the capacities of GfU $g$ and PtG $v$, respectively.

Combining models~(\ref{eqn:GS_full}), (\ref{eqn:SCUC}), coupling constraints~(\ref{eqn:GasPower_coupling}) and the bounds of coupling variables~(\ref{eqn:coupling_var_bound}), the integrated electric-gas system model (IEGS) can be obtained. For brevity, we denote by $\bm{x}$ the binary variables, by $\bm{y}$ the continuous variables, and by $\bm{c}_{\rm{I}}$, $\bm{c}_{\rm{C}}$ the cost vectors associated with them. Eventually, IEGS can be written as,
\begin{subequations} \label{eqn:IEGS}
    \begin{align}
        \nonumber & \underset{\bm{x,y}}{\mathop{\min }}\, \bm{c}_{\rm{I}}^{\top}\bm{x}+\bm{c}_{\rm{C}}^{\top}\bm{y}\\
        \operatorname{s.t.~} & \bm{x} \in \mathcal{X} = \left\{ \bm{x} |\, {\text{(\ref{eqn:SCUC_status})}}-{\text{(\ref{eqn:SCUC_min_down})}} \right\} \\
        \nonumber & \bm{y} \in \mathcal{Y}^{-} = \{\bm{y} |\,  {\text{(\ref{eqn:GS_src_flow_limit})}}-{\text{(\ref{eqn:GS_cmp_lp_f})}}, {\text{(\ref{eqn:GS_f_Pipe_bar_f})}}-{\text{(\ref{eqn:GS_flow_balance})}}, \\
        & ~~~~~~~~~~~~~~~~~~~{\text{(\ref{eqn:SCUC_gen_limit})}}-{\text{(\ref{eqn:SCUC_dc_equation})}}, {\text{(\ref{eqn:GasPower_coupling})}}, {\text{(\ref{eqn:coupling_var_bound})}} \} \\
        & \bm{y} \in \mathcal{Y}^{\rm{GF}} = \left\{ \bm{y} |\,  {\text{(\ref{eqn:GS_f_Pipe_bar_pi})}} \right\}.
    \end{align}
\end{subequations}
The only non-convex part in IEGS is the general flow equaiton~(\ref{eqn:GS_f_Pipe_bar_pi}), which is represented by set~$\mathcal{Y}^{\rm{GF}}$ in Problem~(\ref{eqn:IEGS}).

\subsection{{Uncertainty Modeling}} \label{sec:Incorporating Uncertainties}
To address the variabilities and uncertainties of {renewable energy resources and load demands}, renewable generations as well as electricity/gas demands are viewed as random variables, and a stochastic-programming-based model is developed.
{Conventionally, stochastic programming relies on the probability distribution of random variables. In practice, however, the probability distribution may not exactly exist or the parameters cannot be obtained~\cite{pinson2010conditional}. In recent years, non-parametric statistical methods have been introduced to the power and energy society~\cite{pinson2010conditional, Khorramdel2018Fuzzy}, which help drawing an empirical distribution from historical data without the necessity of assuming any types of distribution for random variables.

In the proposed data-driven method, we first extract the forecast errors from historical data by subtracting the day-ahead forecast values from the real-time values, then use a scenario reduction method to select some representative error scenarios, and finally employ the reduced error scenarios to construct the scenarios by adding the errors to the day-ahead forecast value~\cite{shuai2020real}.
A Wasserstein-metric-based scenario reduction algorithm~\cite{dupavcova2003scenario} is used for scenario reduction. The Wasserstein metric, also known as the Earth Mover's distance, is a function that defines how close two probability distributions are~\cite{liu2018multilevel}, and is more suited for measuring the distance of distributions than many other metrics such as the Euclidean distance. It is worth noting that the reduced scenario set obtained from this algorithm preserves the {correlations} between high-dimensional random variables~\cite{dupavcova2003scenario}.}
\textcolor{black}{It is worth to mention that many other techniques can be introduced to improve the statistical performance of scenario selections. For example, aside from probability metric methods, importance sampling, which aims at selecting scenarios that best represent the average cost impact of uncertainty on the problem~\cite{papavasiliou2013multiarea}, should be a promising alternative.}

In what follows, each realization (scenario) of random nodal injections ${W_{n,t}}$, ${D_{n,t}}$ and ${F^{\rm{Load}}_{s,t}}$ are denoted as $\bm{\xi}$. Moreover, we denote by $\Omega$ the index set of $\bm{\xi}$, {$\bm{\xi}_\omega$ the $\omega$-th scenario, $\bm{y}_\omega$ the $\omega$-th recourse variable, and $\sigma_{\omega}$ the probability of the $\omega$-th scenario.
In two-stage stochastic programming, the second-stage recourse variable is a function of the first-stage decision and the random variable. Therefore, $\mathcal{Y}^{-}$ is written as $\mathcal{Y}^{-}(\bm{x},\bm{\xi})$, and the stochastic integrated electric-gas system model (S-IEGS) can be formulated as follows,}
\begin{subequations} \label{eqn:S-IEGS}
    \begin{align}
     \nonumber & \underset{\bm{x},\bm{y}_\omega} \min\, \bm{c}_{\rm{I}}^{\top}\bm{x} + \sum_{\omega \in \Omega} \sigma_{\omega}\bm{c}_{\rm{C}}^{\top}\bm{y}_{\omega} \\
     \operatorname{s.t.~}& \bm{x} \in \mathcal{X} \\
     & \bm{y}_{\omega} \in \mathcal{Y}^{-}(\bm{x},\bm{\xi}_\omega) \cap \mathcal{Y}^{\rm{GF}} ~\forall \omega\in\Omega,
    \end{align}
\end{subequations}
{in which the on/off statues of CfUs are optimised according to the reduced scenario set, and the second-stage dispatch decision regarding each scenario is determined accordingly.} The price function in S-IEGS is assumed to be in line with that in IEGS. However, it is possible to formulate S-IEGS as a two-settlement process, i.e., attach the pre-dispatch quantity under the forecast scenario with price $\bm{c}_{\rm{C}}$, and multiply the adjusted productions under each scenario with intra-day deviation penalties~\cite{khazaei2017single}.

{The reasons why stochastic programming is preferred in this paper to address the uncertainties in IEGS are threefold:
\begin{enumerate}[1)]
  \item Existing works devoted to stochastic-programming-based co-optimisation problems of IEGS are still limited~\cite{alabdulwahab2015coordination, mirzaei2020novel}.
  \item As shown in Section~\ref{sec:alg} and \ref{sec:case_1}, taking the advantage of stochastic programming, the solution procedure ends up iteratively solving some {separable} convex problems, the convergence and optimality of which are guaranteed.
  \item Although cutting-edge techniques like (distributionally) robust optimisation can also deal with uncertainties, they make the MINLP problem rather complicated, so that approximation algorithms (not only for the physical model itself)~\cite{wu2019distributionally} and heuristics~\cite{he2016robust} become inevitable.
\end{enumerate}

To support the viewpoints above, distributionally robust optimisation is adopted for comparison. The distributionally robust integrated electric-gas system model (DR-IEGS) can be formulated as follows,
\begin{subequations} \label{eqn:DR-IEGS}
    \begin{align}
        \nonumber & \underset{\bm{x}}{\mathop{\min }}\, \bm{c}_{\rm{I}}^{\top}\bm{x} + \max_{\mathbb{P}\in\mathcal{P}}\, \min_{\bm{y}}\, \mathbb{E}_{\mathbb{P}}[\bm{c}_{\rm{C}}^{\top}\bm{y}]\\
        \operatorname{s.t.~} & \bm{x} \in \mathcal{X} \\
        & \bm{y} \in \mathcal{Y}^{-} \cap \mathcal{Y}^{\rm{GF}} \\
        & \mathcal{P} \in \mathcal{P}_0(\Xi),
    \end{align}
\end{subequations}
where $\Xi$ is the feasible region of $\bm{\xi}$, ${{\mathcal{P}}}_{0}(\Xi)$ denotes the set of all probability measures on a sigma algebra of $\Xi$, and the subset $\mathcal{P}$ is known as the ambiguity set in distributionally robust optimisation~\cite{wu2019distributionally, fang2019introducing}.
For tractability, only linear moment constraints are considered in the ambiguity set as in Ref.~\cite{xiong2017distributionally, wu2019distributionally}. It should be noted that linear moment is not capable of modeling the correlation of uncertainties.

The distributionally robust model~(\ref{eqn:DR-IEGS}) is also a data-driven approach. Historical data is used to construct the ambiguity set, among which the model aims to seek a \emph{worst-case} distribution. The main difference between S-IEGS and DR-IEGS is that the optimal decision is derived based on the generated scenarios in Problem~(\ref{eqn:S-IEGS}), whereas the optimal decision is achieved regarding the worst-case distribution in Problem~(\ref{eqn:DR-IEGS}).

Assuming that historical data is available to both S-IEGS and DR-IEGS, one can show that the stochastic model has the advantage over the distributionally robust model in terms of tractability and in-sample/out-of-sample performances. These will be demonstrated in Section~\ref{sec:alg} and Section~\ref{sec:case}.
}

\subsection{Pricing PtGs in Day-ahead Market under Uncertainties} \label{sec:IEGS_pricing}
The main role that PtGs play in the integrated system is to consume surplus renewable generations and produce natural gas. Therefore, the contributions of PtGs are twofold: \textit{i}) reducing the penalty cost (or the environmental cost) of renewable generation curtailments, and \textit{ii}) supplying additional natural gas. It is necessary to quantify such contributions, especially in a competitive market. One common method is using the locational marginal prices (LMPs){, which are the optimal Lagrangian multipliers of the optimisation problem that determine the costs of producing one extra unit of resource at different loccations~\cite{gomez2008electric}}.
If we associate with the gas balance equation~(\ref{eqn:GS_flow_balance}) and the power flow equation~(\ref{eqn:SCUC_dc_equation}) Lagrangian multipliers $\bm{\lambda}$ and $\bm{\mu}$ respectively, then the ``net'' LMP (or LMP simply) of PtG $v$ that defined in \$/MW from the PtG's perspective is
\begin{equation}\label{eqn:LMP_PtG}
  \psi_{v,t} = \mu_{m,t|\kappa^{\rm{PtG}}_{v,m}=1} - \eta^{\rm{PtG}}_{v}/H_{\rm{g}}\lambda_{n,t|{v\in\mathcal{G}^{\rm{PtG}}_n}},
\end{equation}
where $\lambda_{n,t}$ is the multiplier of (\ref{eqn:GS_flow_balance}) for node $n$ at time $t$, and $\mu_{m,t}$ is the multiplier of (\ref{eqn:SCUC_dc_equation}) for bus $m$ at time $t$. Both $\lambda_{n,t}$ and $\mu_{n,t}$ can be either positive, zero, or negative.

It can be proved that when Problem~(\ref{eqn:IEGS}) is solved to optimality~\textcolor{black}{[Since Problem~\eqref{eqn:IEGS} is a MINLP, solving it to optimality is defined herein as: fixing the binary variables as their optima, and re-solving the NLP problem to optimality (maybe local optimality) to obtain the optimal multipliers.]}: \textit{i}) the electric power consumed by PtG $v$ is non-zero if and only if $\psi_{v,t}$ is \emph{non-positive}; \textit{ii}) $\psi_{v,t}$ is \emph{negative} if and only if the capacity of PtG is inadequate. The former holds since otherwise the conversion would increase the total cost{. The latter holds since otherwise the PtG production level can be improved to further reduce the total cost, which is contradictory with the fact that the current solution is optimal. The second observation} suggests that PtG can only profit from congestion under the LMP-based pricing regime.

Evidently, the above-mentioned LMP only reflects the marginal value of PtG under {a certain scenario (i.e., the forecast scenario), and it doesn't accounts for the flexibility service that PtG could provide after the realization of uncertainty.} Due to the significant randomness in day-ahead markets, it is crucial to price the flexible resources provided by PtGs that mitigate the uncertainties~\cite{ye2016uncertainty, fang2019introducing}. As such, a novel concept of expected locational marginal value (E-LMV) is proposed in this paper. E-LMV can be formed with the byproduct of solving S-IEGS:
\begin{equation}\label{eqn:E-LMV_PtG}
   \mathbb{E}[{\Psi}_{v,t}] \overset{\vartriangle}= -\sum_{\omega\in\Omega} \sigma_{\omega} \psi_{v,t,\omega} p^{\rm{PtG}}_{v,t,\omega},
\end{equation}
where the subscript $\omega$ of $\psi_{v,t,\omega}$ {and $p^{\rm{PtG}}_{v,t,\omega}$ indicates that they are} derived from the $\omega$-th scenario. Intuitively, E-LMV represents the expectation of payment that is entitled to PtG, regarding its potential recourse actions after uncertainties reveal. {By taking the expectation value of multiple LMPs, E-LMV provides a payment scheme that is closer to the ``true'' (in terms of mathematical expectation) LMP, and therefore is suited for a market with considerable uncertainties.}

E-LMVs can be defined similarly for the other participants in the day-ahead market.
{For example, E-LMV of RES at bus $n$ at time $t$ is given by
\begin{equation}\label{eqn:E-LMV_RES}
   \mathbb{E}[{\Psi}^{\rm{RES}}_{n,t}] \overset{\vartriangle}= \sum_{\omega\in\Omega} \sigma_{\omega}\mu_{n,t,\omega}W_{n,t,\omega}.
\end{equation}
}

Ultimately, the day-ahead market {is settled} based on E-LMVs. {We have the following proposition for E-LMVs (see Appendix for the proof and further discussions)}, which suggests that the money collected by system operators from consumers is more than that should be paid to suppliers.

{
\begin{proposition} \label{pro:1}
Supposing there is no gas compressors in the natural gas system, E-LMVs ensure revenue adequacy for the integrated electric-gas system.
\end{proposition}
}

{For the distributionally robust model, we propose using the extremal distribution $\mathbb{P}^{\ast}$ to derive E-LMV. Since a distributionally robust optimisation problem always possesses a discrete extremal distribution, $\mathbb{E}_{\mathbb{P}^{\ast}}[{\Psi}_{v,t}]$ can be calculated using the extremal distribution as in Eqn.~(\ref{eqn:E-LMV_PtG}). Intuitively, E-LMV yielded from the distributionally robust model should be higher than that from the stochastic model; this will be verified in Section~\ref{sec:case}.
}

\section{Solution Algorithm}\label{sec:alg}
In this section, we first introduce a method to address the non-convexity issue for the natural gas flow model, and then present the overall solution algorithm for S-IEGS.
\subsection{Convexification of Nonlinear General Flow Equation}
The most challenging part of Problem~(\ref{eqn:S-IEGS}) is the non-convexity of the general flow equation, as detailed in Eqn.~(\ref{eqn:GS_f_Pipe_bar_pi}).
\textcolor{black}{Techniques for tackling this difficulty can be divided into: $i$) nonlinear programming (NLP) methods that solve the problem with interior point methods, etc.; $ii$) MILP reformulation and second-order cone programming (SOCP) approximation~\cite{chen2019unit,wang2018risk} that aim to approximate with high accuracy the non-convex problem using tractable mathematical programmings; and $iii$) intelligent algorithms like particle swarm optimisation, genetic algorithm, and neural networks~\cite{baghaee20161reliability, parizad2020security, baghaee2016power, baghaee2017fuzzy}. Noting that SOCP approximation enjoys higher computational efficiency, and it is such that decomposition methods could be easily implemented, we adopt it in this paper.}

Assuming that the direction of gas flow is known~\cite{chen2019unit, he2016robust}, then the general flow equation~(\ref{eqn:GS_f_Pipe_bar_pi}) can be written as,
\begin{equation} \label{eqn:general_flow}
\left\{ \begin{aligned}
        & K^{\rm{gf}^2}_{(m,n)} \pi^2_{m,t} \ge \bar{f}^{\rm{Pipe}^2}_{(m,n),t} + K^{\rm{gf}^2}_{(m,n)}\pi^2_{n,t}  \\
        & K^{\rm{gf}^2}_{(m,n)} \pi^2_{m,t} - K^{\rm{gf}^2}_{(m,n)}\pi^2_{n,t} - \bar{f}^{\rm{Pipe}^2}_{(m,n),t} \le 0. \\
    \end{aligned} \right.
\end{equation}
The first row in Eqn.~(\ref{eqn:general_flow}) defines a second-order cone:
\begin{equation*}\label{eqn:cone1}
\small
\begin{aligned}
  \mathcal{Q}^1_{(m,n),t} = \left\{ \left(\tilde{\pi}_{m,t}, \bar{f}^{\rm{Pipe}}_{(m,n),t}, \tilde{\pi}_{n,t}\right) \Big|\,
   \tilde{\pi}_{m,t} \ge \left\| \left[\bar{f}^{\rm{Pipe}}_{(m,n),t}; \tilde{\pi}_{n,t}\right] \right\| \right\},
  \end{aligned}
\end{equation*}
where $\tilde{\pi}_{m,t} = K^{\rm{gf}}_{(m,n)}\pi_{m,t}$ and $\tilde{\pi}_{n,t} = K^{\rm{gf}}_{(m,n)}\pi_{n,t}$.
The second row in Eqn.~(\ref{eqn:general_flow}) results in a DC (difference of convex functions) programming that is difficult to solve in general. According to~\cite{lipp2016variations}, DC programming can be approximately solved by a penalty convex-concave procedure (PCC). Specifically, the concave items are linearised at the current points, yielding a convex problem (SOCP in this paper; hence a sequential SOCP method), and then sequentially, the convex problem is solved to update the points for linearization. To ensure feasibility, a positive slack variable is needed:
\begin{equation*}\label{eqn:linearization}
\small
\begin{aligned}
    \mathcal{Q}^2_{(m,n),t} = \Big\{ & \left(\tilde{\pi}_{m,t}, \bar{f}^{\rm{Pipe}}_{(m,n),t}, \tilde{\pi}_{n,t}\right) \Big|\, s^{+}_{(m,n),t} \ge 0,\\
    & K^{\rm{gf}^2}_{(m,n)} \pi^2_{m,t} - K^{\rm{gf}^2}_{(m,n)} \left(2\pi^{\ast}_{n,t}\pi_{n,t} - \pi^{\ast^2}_{n,t} \right) - \\
    & \left( 2\bar{f}^{\rm{Pipe}^{\ast}}_{(m,n),t}\bar{f}^{\rm{Pipe}}_{(m,n),t} - \bar{f}^{\rm{Pipe}^{\ast2}}_{(m,n),t} \right) \le s^{+}_{(m,n),t} \Big\}. \\
  \end{aligned}
\end{equation*}
The intersection of $\mathcal{Q}^1_{(m,n),t}$ and $\mathcal{Q}^2_{(m,n),t}$ equivalently forms the feasible set of constraint~(\ref{eqn:general_flow}) only if $s^{+}_{(m,n),t}$ vanishes.

For brevity, we define for each scenario the convex approximation of $\mathcal{Y}^{\rm{GF}}$ as $\mathcal{Q}$, which is given by
\begin{equation*}\label{eqn:linearization}
    \begin{aligned}
        \mathcal{Q} = \Big\{\bm{y} \Big| & \left(\tilde{\pi}_{m,t}, \bar{f}^{\rm{Pipe}}_{(m,n),t}, \tilde{\pi}_{n,t}\right) \in \\
        & \mathcal{Q}^1_{(m,n),t} \cap \mathcal{Q}^2_{(m,n),t} ~\forall (m,n)\in\mathcal{G}^{\rm{Pipe}}, t\in\mathcal{T} \Big\}.
  \end{aligned}
\end{equation*}
Moreover, for ease of exposition, {a normalised slack variable is defined as $\tilde{\bm{s}}^{+}$}, the $[{(m,n),t}]\text{-th}$ entry of which is given by $s^{+}_{(m,n),t} \slash (K^{\rm{gf}^2}_{(m,n)} \pi^2_{m,t})$.

\subsection{Generalised Benders Decomposition with PCC}
Incorporating PCC into the generalised Benders decomposition procedure, an algorithm for solving S-IEGS can be obtained, as detailed in Algorithm~\ref{alg:1}. The Benders sub-problem is modified to avoid the necessity of solving a dual SOCP problem. Specifically, by introducing equality constraint~(\ref{eqn:BD_sub_x_fix}), it can be proved that the optimal dual variable associated with this constraint, which is available from off-the-shelf solvers, is sufficient to construct a Benders cut. Besides, in order for a valid cut, \emph{strong duality} must hold for the Benders sub-problem, which in turn requires that Problem~(\ref{eqn:BD_sub}) and its dual have strictly feasible solutions, i.e., $\mathcal{Q}^1_{(m,n),t}$ and $\mathcal{Q}^2_{(m,n),t}$ have non-empty interior~\cite{alizadeh2003second}. In computational practice, the feasibility condition is ensured by introducing slack variables to the power/gas balance equations (and penalty costs to the objective function accordingly), while the non-empty interior condition is guaranteed by the slack variable of $\mathcal{Q}^2_{(m,n),t}$.
\begin{algorithm}[t]
\caption{Generalised Benders Decomposition with PCC}
\label{alg:1}
\small
\begin{algorithmic}[1]
    \State Select convergence tolerance $\epsilon$, $\varepsilon$ and $\delta$; select initial/maximum penalty factor $\rho / \overline{\rho}$ and $\varsigma > 1$; denote the optimum of $x$ as $x^{\ast}$; let $LB = -\infty$, $UB$ and $\widehat{UB} = \infty$; let $i=0$, $\mathcal{J}=\varnothing$.
    \While{$(UB-LB)/UB \ge \epsilon$}
        \State Solve the Benders master problem~(\ref{eqn:BD_master})
        \begin{subequations} \label{eqn:BD_master}
            \begin{align}
                 \nonumber & \underset{\bm{x},\gamma} \min\, \bm{c}_{\rm{I}}^{\top}\bm{x} + \gamma \\
                 \operatorname{s.t.~}& \bm{x} \in \mathcal{X}, ~\gamma \ge 0 \\
                 & \gamma \ge V(\bm{x}_{j}) + \bm{\nu}_{j}^{\top}(\bm{x}-\bm{x}_{j}) ~\forall j \in \mathcal{J} \label{eqn:BD_master_cut}
            \end{align}
        \end{subequations}
        \State $i \gets i+1$, $\mathcal{J} \gets \mathcal{J} \cup \{i\}$, $\bm{x}_{i} \gets \bm{x}^{\ast}$, $LB \gets  \bm{c}_{\rm{I}}^{\top}\bm{x}^{\ast} + \gamma^{\ast}$
        \Repeat 
            \State Solve the current approximation, Problem~(\ref{eqn:BD_sub})
            \begin{subequations} \label{eqn:BD_sub}
            \begin{align}
                 \nonumber V(\bm{x}_{i}) = & \underset{\bm{y}_{\omega},\bm{s}^{+}_{\omega},\bm{z}} \min\, \sum_{\omega\in\Omega} \sigma_{\omega}(\bm{c}_{\rm{C}}^{\top}\bm{y}_{\omega} + \rho\bm{1}^{\top}\bm{s}^{+}_{\omega}) \\
                 \operatorname{s.t.~}& \bm{z} = \bm{x}_{i} ~: \bm{\nu} \label{eqn:BD_sub_x_fix} \\
                 & \bm{y}_{\omega} \in \mathcal{Y}^{-}(\bm{x}_{i},\bm{\xi}_\omega) \cap \mathcal{Q} ~\forall \omega\in\Omega
            \end{align}
        \end{subequations}
        \State $\rho \gets \min\{\varsigma\rho,\overline{\rho}\}$, $\widehat{UB} \gets UB$, $UB \gets V(\bm{x}_{i})$, $\bm{\nu}_{i} \gets \bm{\nu}^{\ast}$
        \Until $|UB-\widehat{UB}|/UB \le \varepsilon$ and $\|\tilde{\bm{s}}^{+^{\ast}}_{\omega}\|_{\infty} \le \delta ~\forall \omega\in\Omega$
    \EndWhile
    \State return $\bm{x}_{i}$
\end{algorithmic}
\end{algorithm}

The proposed algorithm has some desirable properties:
\begin{enumerate}[1)]
  \item It is separable with regard to each scenario, and hence Problem~(\ref{eqn:BD_sub}) can be solved in parallel.
  \item Problem~(\ref{eqn:BD_sub}) or its separated problems are convex, and thus can be solved in a decentralised manner by the electric system operator and the natural gas system operator, e.g., using ADMM.
  \item If at each Benders iteration, Problem~(\ref{eqn:BD_sub}) can be distributedly solved to optimality, since MILP~(\ref{eqn:BD_master}) involves merely the electric power system model, then, without confidential information of each system being revealed, Algorithm~\ref{alg:1} converges and returns a UC solution.
\end{enumerate}

{\color{black}
\begin{figure}[t]
  \centering
  \color{black}
  \includegraphics[width=2.8in]{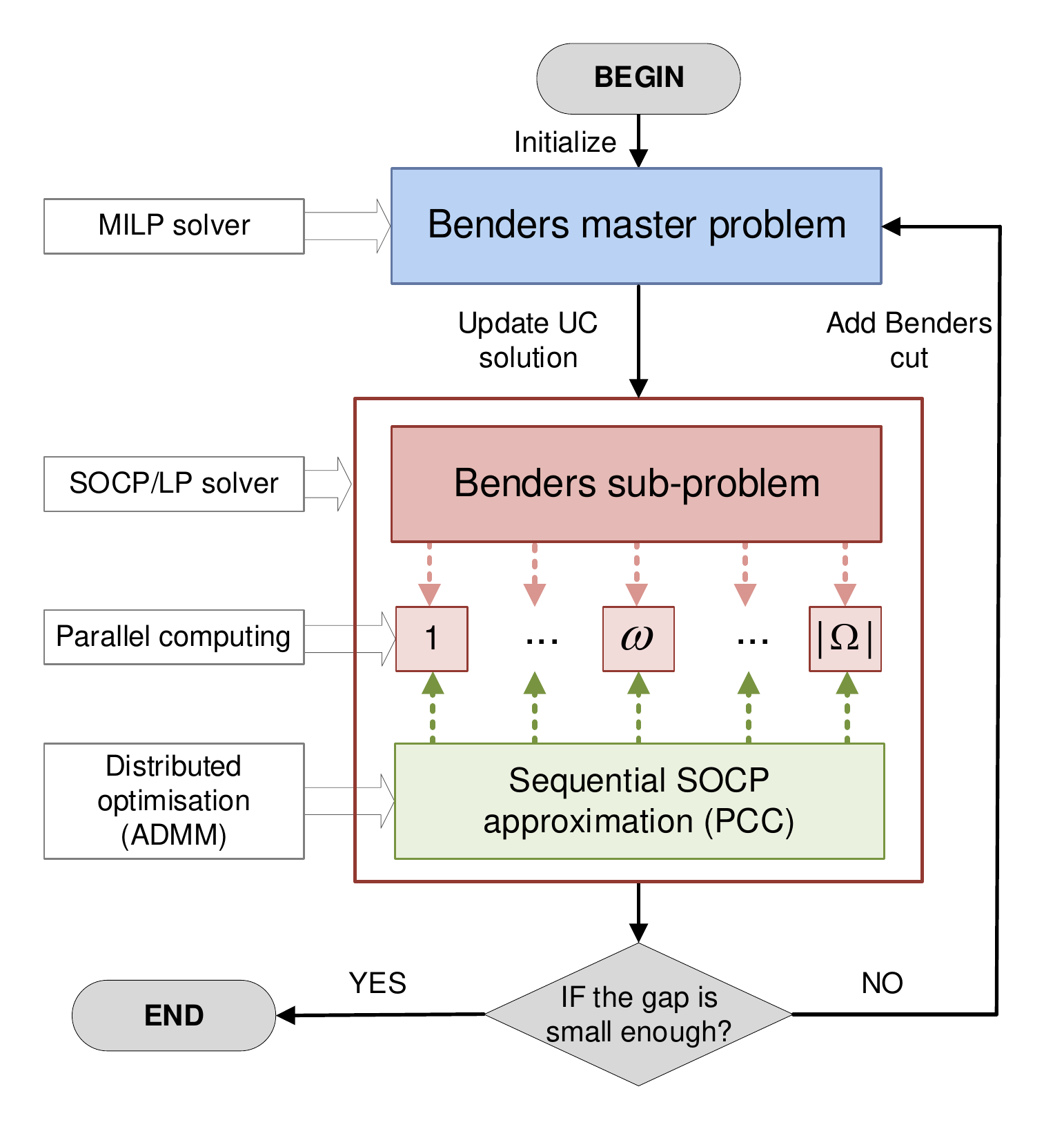}
  \caption{\color{black}Flowchart of the solution algorithm.}\label{fig:alg}
\end{figure}
For ease of reading, the framework of the whole solution algorithm is provided, which is shown in Fig.~\ref{fig:alg}. The outer loop of the algorithm is the generalised Benders decomposition that iterates from the MILP master problem and the convex sub-problem. The Benders sub-problem is parallelizable, which means $|\Omega|$ scenarios could be addressed with PCC meanwhile. As mentioned above, the convex sub-problems can be decomposed into a linear programming (LP) of the power system dispatch problem and an SOCP of the gas flow problem, and then coordinated with ADMM.
}

{
\subsection{Solution Method for Distributionally Robust Model}
For comparison purpose, the distributionally robust model will also be solved. Yet, the convexification method and PCC algorithm cannot be easily extended to the distributionally robust model. One reason is that the convexified model is nonlinear, and thus the state-of-the-art method, linear decision rule (LDR) is inapplicable~\cite{xiong2017distributionally, wu2019distributionally}. Another obstacle is that if we choose fully adaptive recourse instead of LDR, then the solution procedure requires dualizing the second-stage problem, making it unclear how to sequentially penalize the primal constraints.

To this end, Taylor series expansion is applied to linearize Eqn.~(\ref{eqn:GS_f_Pipe_bar_pi}) for the distributionally robust model~\cite{wu2019distributionally}. Although the linearised model is favorable for developing solution algorithm, it is less tight than the sequential SOCP method. Due to the above-mentioned limitations, it is recognised that distributionally robust optimisation is not so attractive to the already complicated non-convex IEGS problem.

The distributionally robust model is solved by an extremal distribution generation method proposed in~\cite{zheng2020data}. The by-product of the solution method is an extremal distribution, which is then used for E-LMV calculation.}

\section{Case Studies} \label{sec:case}
In this section, numerical experiments are carried out to validate the effectiveness of {the proposed stochastic-programming-based model}, the efficiency of the proposed solution method, and the advantage of the pricing method.
\begin{figure*}[!b]
  \centering
  \includegraphics[width=5.8in]{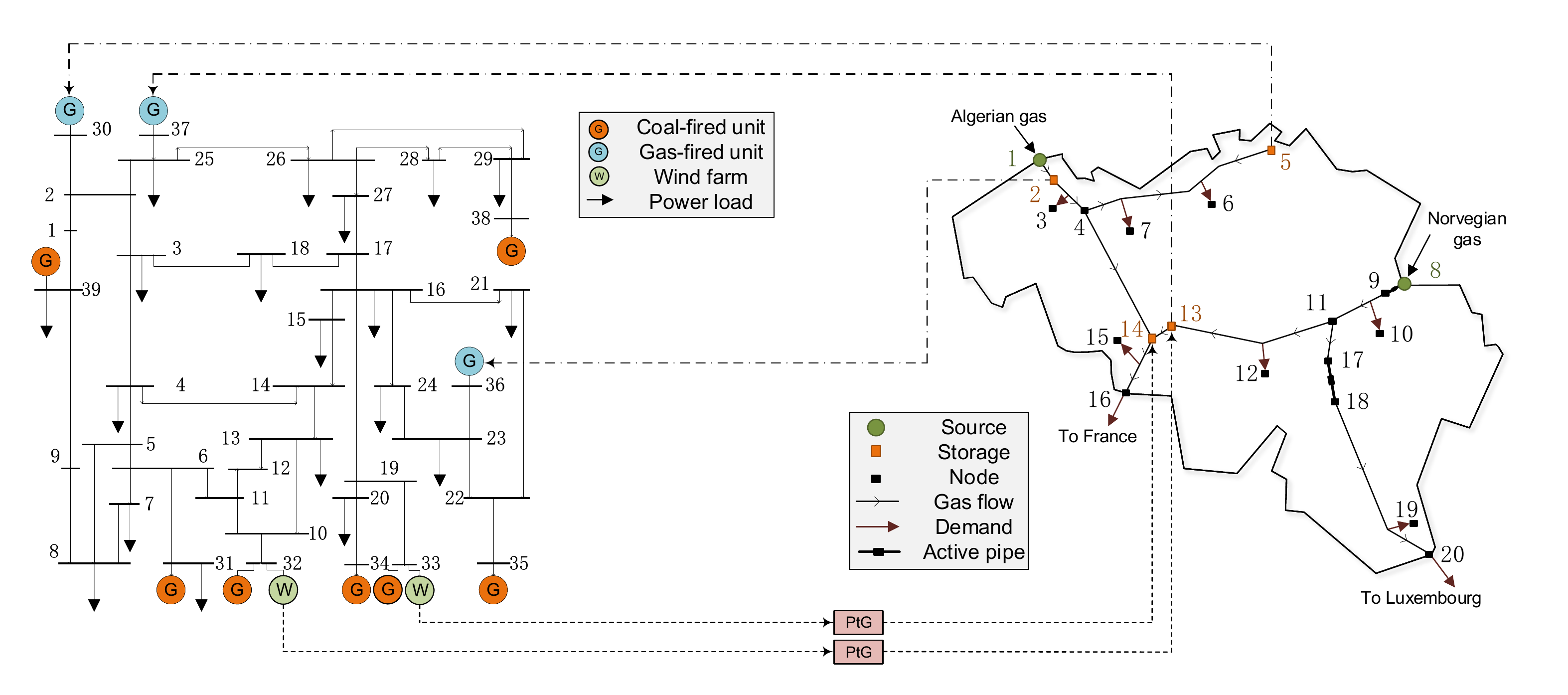}
  \caption{Diagram of integrated electricity-gas system (IEEE 39-bus system and Belgium 20-node gas system).}\label{fig:39-bus-20-node}
\end{figure*}

The test system is obtained by combining the IEEE 39-bus system and the Belgium 20-node gas system. The configuration of the integrated system is exactly as shown in Fig.~\ref{fig:39-bus-20-node}~\cite{wei2017power}, and detailed data is available online~\cite{zheng_2019}. Two 1200-MW wind farms are located at Bus 32 and Bus 33, resulting in a wind power penetration rate of 24.6\%.
\textcolor{black}{In order to hedge against the volatile wind power generation and help consuming extra wind power, two 200-MW PtGs are installed near the wind farms, and the gas is injected into Node 13 and Node 14 of the gas system, respectively. The GfUs located at Bus 30, Bus 36, and Bus 37 are supplied by the gas extracted from Node 5, Node 2, and Node 13, respectively.}

The day-ahead {forecast} and real-time data series of wind farm outputs and load demands over one year are adopted~\cite{pena2017extended}. After scaling, we generate error scenarios with 85\% of the data series (the day-ahead forecast errors of wind power and load demands are assumed to be $\pm$50\% and $\pm$10\% respectively), and randomly remain 15\% of them for out-of-sample tests.
\textcolor{black}{According to current practise, the penalty costs of wind curtailment and electric/gas load shedding are set higher in order to mimic the environmental cost, and reduce the loss of load, respectively. Without loss of generality, in the case studies, the price of wind curtailment is set to 10 times of the mean cost of power generation in the test system, namely 142 \$/MWh; the prices of electric load shedding and gas load shedding are set to 200 times of the mean cost of power generation and the mean gas price in the test systems, namely 2840 \$/MWh and 396 \$/MBTU, respectively.
}

The optimisation problems are built in GAMS 26.1.0 and solved by CPLEX 12.8. The relative convergence tolerance of CPLEX and those in Algorihtm~\ref{alg:1} are all set as 10\textsuperscript{-4}. All runs are executed on an Intel i5 CPU machine running at 1.80 GHz with 8 GB of RAM.

\subsection{Performances of Proposed Algorithm} \label{sec:case_1}
\begin{figure}[t]
  \centering
  \includegraphics[width=3.3in]{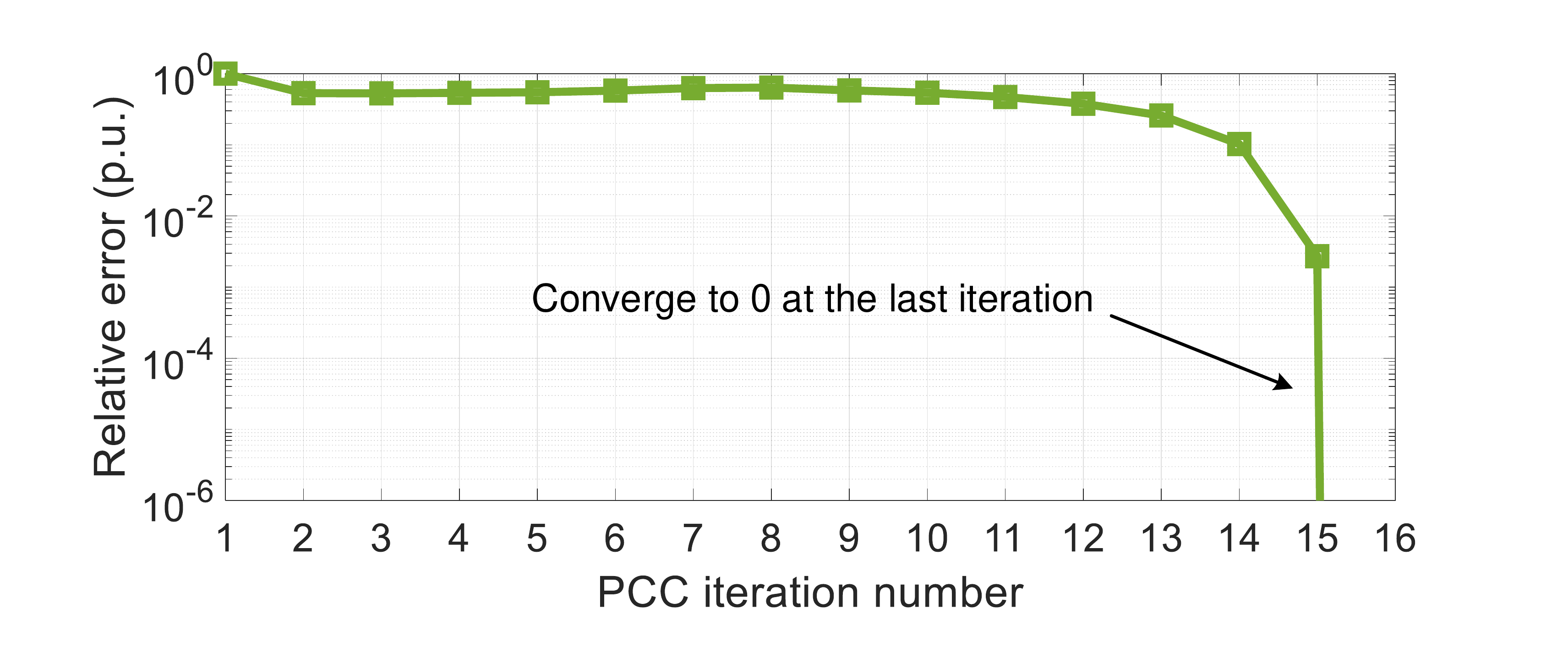}
  \caption{Relative violation of general flow equation at each PCC iteration.}\label{fig:SOCP_cons_error}
\end{figure}

The efficiencies of the proposed algorithm is verified on multiple cases. The Benders loop converges with predefined accuracy (i.e., 10\textsuperscript{-4}, and it converges to {a} zero gap in some cases) after 42 to 66 iterations. The PCC loop takes about 16 iterations, and the slack variables in $\mathcal{Q}^2_{(m,n),t}$ usually vanish (see Fig.~\ref{fig:SOCP_cons_error}), indicating that the solution is feasible to the primal MINLP.
\textcolor{black}{Despite being less computationally expensive, the linearised model used by DR-IEGS always produces non-zero residuals of the relaxed gas flow equations. So DR-IEGS seldom achieves a feasible solution to the primal MINLP, as also reported in~\cite{wu2019distributionally}.}

The accuracy of Algorithm~\ref{alg:1} is demonstrated via Table~\ref{tab:accuracy_runtime}. For the nonlinear gas model, PCC finds a solution extremely close to the one returned by IPOPT, albeit it becomes more time-consuming due to a smaller step-size of the penalty factor (i.e., $\varsigma$=1.02). For IEGS, Algorithm~\ref{alg:1} finds a solution that is only 0.061\% larger than the feasible solution returned by COUENNE, which exhaustedly runs out of time.
\begin{table}[t]
\scriptsize
\centering
\caption{Accuracy and Runtime Compared with (MI-)NLP Solvers}\label{tab:accuracy_runtime}
\begin{tabular}{l|cccc}
\hline
Model                 & \multicolumn{2}{c}{Gas system model}                   & \multicolumn{2}{c}{IEGS}             \\ \hline
Formulation           & NLP      & \multicolumn{1}{c|}{SOCP ($\varsigma$=1.02)}     & MINLP    & MISOCP ($\varsigma$=2)                    \\
Obj. (k\$)             & 1,714.93 & \multicolumn{1}{c|}{1,714.94} & {3,069.20} & {3,071.06}                  \\
Runtime (s)           & 54.53    & \multicolumn{1}{c|}{92.37}    & {36000.00}    & {1039.03} \\
Error (p.u.) & \multicolumn{2}{c|}{0.001\%}             & \multicolumn{2}{c}{0.061\%}          \\ \hline
\end{tabular}
\end{table}

The total computational time of solving S-IEGS is reported in Table~\ref{tab:ite_runtime}. Since the Benders sub-problem is separable, when leveraging parallel computations, the algorithm can actually terminate within 30 minutes even for the 100-scenario case {(the average runtime of each scenario ranges from 952.57 seconds to 1727.08 seconds)}, thus meeting the time requirement of day-ahead markets.
\textcolor{black}{In order to test the scalability of the proposed algorithm, we replace the 39-bus system with the IEEE 118-bus system. Numerical results show that if we only impose power flow limits on critical transmission lines instead of all lines as in engineering practise, the S-IEGS problem is solvable within 2 hours accounting for the effect of parallel computation. Specifically, the relative gap of the Benders loop could be closed to about 10\textsuperscript{-3} within 100 iterations, and PCC basically converges within 20 iterations. Although the number of iterations needed to solve S-IEGS is about 20 times (i.e., the average number PCC iterations) of that needed to solve a stochastic UC problem with similar scale, the overall computational effort turns out to be acceptable as the SOCPs could be solved quite efficiently.}

\begin{figure}[t]
  \centering

  \includegraphics[width=3.3in]{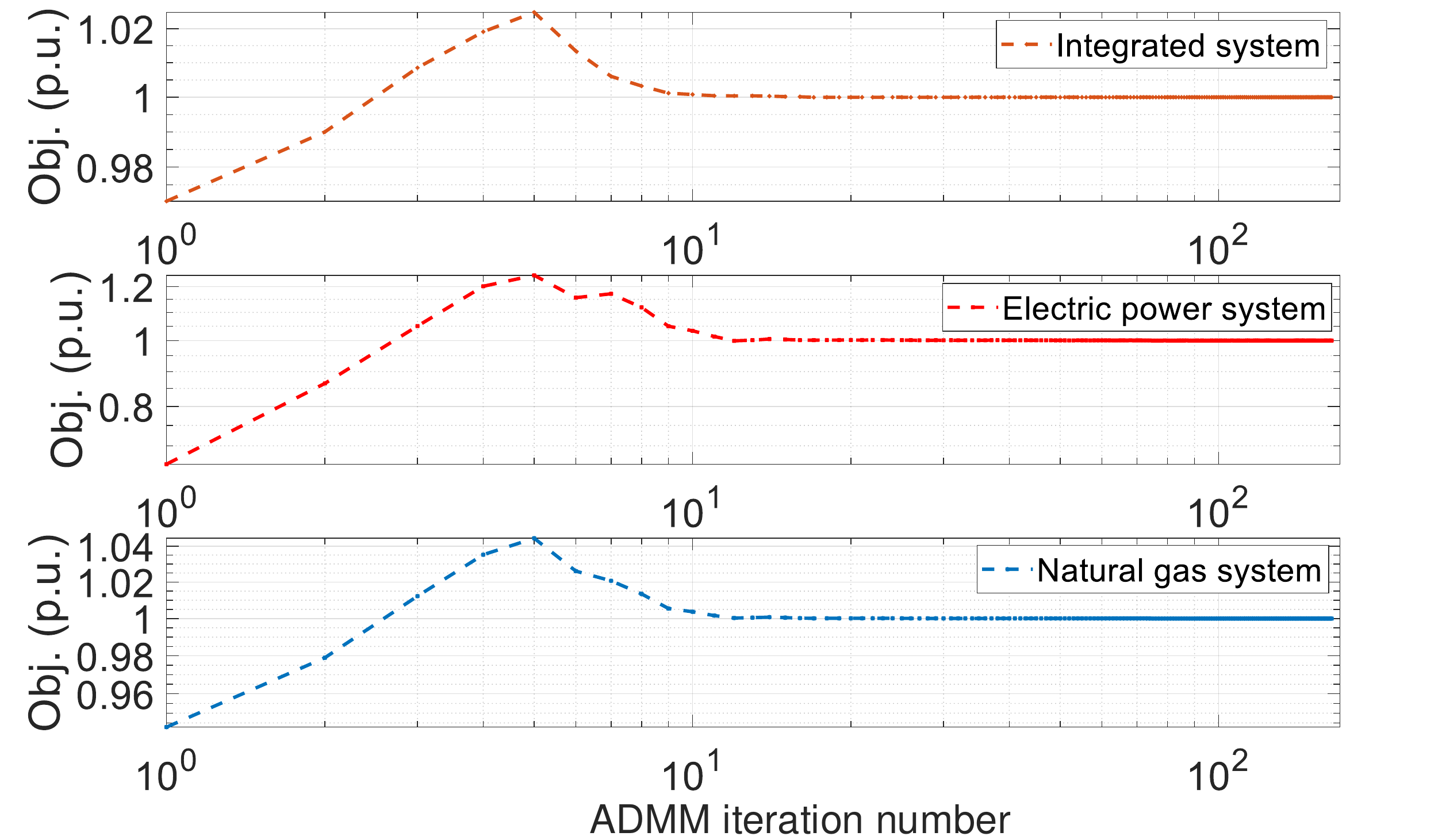}\\
  \caption{Convergence profile of ADMM for Benders sub-problem~(\ref{eqn:BD_sub}).}\label{fig:ADMM_socp_convergence}
\end{figure}

We also deploy { the standard ADMM~\cite{boyd2011distributed}} to Problem~(\ref{eqn:BD_sub}), and find that the two-block SOCP can be solved to global optimality within 200 iterations{, or solved to a 10\textsuperscript{-4} gap within 20 iterations (see Fig.~\ref{fig:ADMM_socp_convergence}). The runtime of ADMM for the test system is several minutes. It is worth mentioning that in DR-IEGS, the sub-problem cannot be decomposed and precisely solved by the electric system operator and the natural gas system operator.}
\begin{table}[t]
\footnotesize

\centering
\caption{Computational Efficiencies of Proposed Algorithm} \label{tab:ite_runtime}
\begin{tabular}{c|ccc}
\hline
\# of scenarios & \# of Benders iter.   & \# of PCC iter.       & Runtime (sec.)      \\ \hline
1        & 42         & 16         & 1039.03        \\
5         & 43         & 15-16      & 4762.85       \\
10        & 45         & 15-16      & 10481.82      \\
20        & 43         & 15-16      & 20326.36      \\
50        & 50         & 15-16      & 61062.44      \\
100       & 66         & 15-17      & 172708.17     \\ \hline
\end{tabular}
\end{table}

{Therefore, the stochastic model and the proposed solution method is practicable and favorable in terms of efficiency, accuracy, scalability and the possibility of distributed computing.
}

\subsection{{Advantages of Proposed Stochastic Method}}
In this subsection, the improvement of UC decision brought by stochastic programming is evaluated. {The benchmarks include a deterministic IEGS model (D-IEGS), which deals with uncertainties by operational reserves (the reserve rates for the gas system and the power system are 5\% and 10\% respectively), and the distributionally robust model described in Section~\ref{sec:Incorporating Uncertainties}}.

\begin{figure*}[!tb]
  \centering
  \color{black}
  \includegraphics[width=6in]{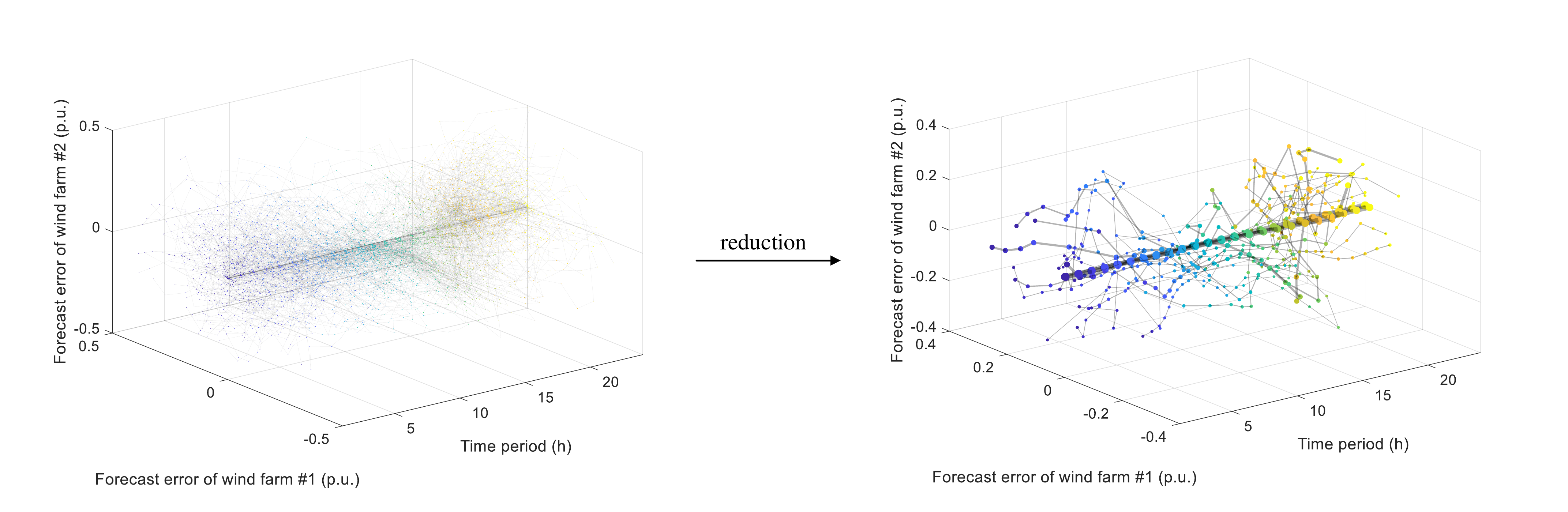}
  \caption{\color{black}Scenario reduction for wind power forecast errors.\label{fig:reduced_scenarios}}
\end{figure*}

\textcolor{black}{The scenario reduction process is illustrated by Fig.~\ref{fig:reduced_scenarios}.
The left panel of Fig.~\ref{fig:reduced_scenarios} shows the 312 historical observations of wind power forecast error of two wind farms; the right panel of Fig.~\ref{fig:reduced_scenarios} shows the 20 reduced scenarios, in which a scenario with higher probability is plotted with a heavier line.}
By using the algorithm in~\cite{liu2018multilevel}, the Wasserstein distance between the reduced scenario set and the original data can be approximated. As shown in Table~\ref{tab:UC_to_scenarios}, the asymptotics of the reduced scenario sets is quite obvious, i.e., the distribution gets closer to the empirical one as the scenario size grows.
As expected, the UC solution varies with the scenario size, and it ``converges'' as the number of scenarios becomes sufficiently large (see Table~\ref{tab:UC_to_scenarios}). In fact, only two ``sub-optimal'' UC solutions occur, which have distinct on/off statues over 10 or 1 time slots compared with the ``optimal'' one. We find that 20 scenarios might be representative enough for this case. {It is observed that the extremal distribution yielded from DR-IEGS is quite ``far'' from the empirical distribution, and the UC solution also differs a lot with those of S-IEGS.}

\begin{table}[t]
\footnotesize
\small
\center

\caption{Comparisons of Probability Distributions and UC Solutions\\under Different Methods and Scenario Sizes}\label{tab:UC_to_scenarios}
\begin{tabular}{c|cc}
\hline
Method/ & Distance to empirical& \# of distinct  \\
\# of scenario &  distribution (p.u.) & on/off statues (h) \\ \hline
D-IEGS & 2.86 & 10 \\
5 scenarios & 2.46 & 10 \\
10 scenarios & 2.14 & 10 \\
20 scenarios & 1.91 & 1 \\
50 scenarios & 1.28 & 1 \\
100 scenarios & 1.00 & 0 \\
DR-IEGS & 16.55 & 38 \\ \hline
\end{tabular}
\end{table}

\begin{figure}[t]
  \centering
  \includegraphics[width=3.4in]{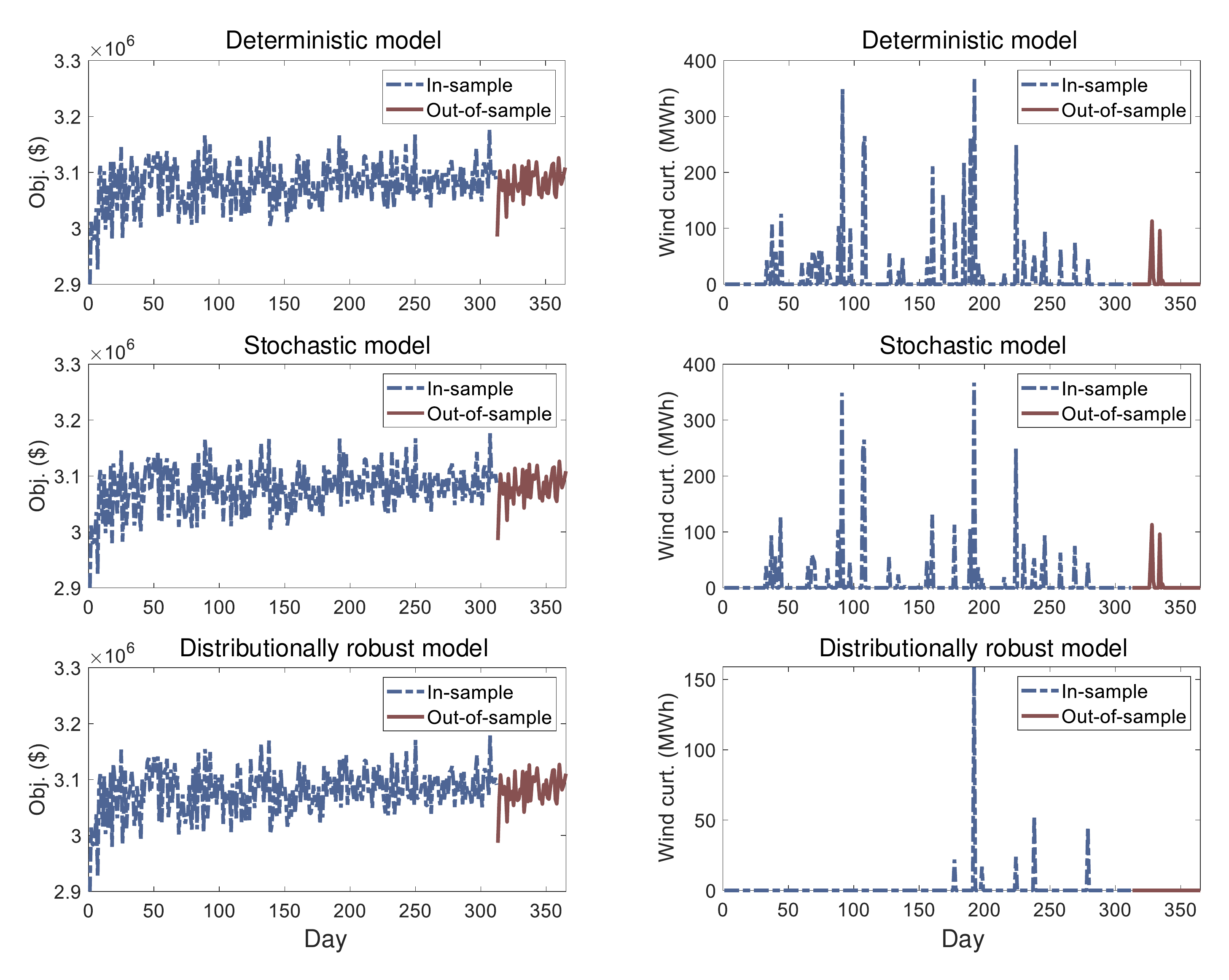}
  \caption{Simulation results of deterministic model, stochastic model and distributionally robust model.}\label{fig:Cost_and_curt}
\end{figure}

For all the methods, after a UC decision is derived, in-sample and out-of-sample simulations are carried out to yield the expected costs under this UC solution. The simulation results for all the methods are presented in Fig.~\ref{fig:Cost_and_curt}. The stochastic model outperforms the deterministic one slightly in terms of the amount of wind curtailments. As shown in Table~\ref{tab:cost_and_curt}, although the stochastic model incurs wind curtailments in the scheduling phase, the UC solution derived from it does reduce 2.17-MWh wind curtailments in simulations. \textcolor{black}{Thus, the proposed method facilitates the utilization of wind power more effectively than the comparative decision making methods, and helps reducing the impact of greenhouse gas emission better.}
The cost saving achieved by optimizing the UC decision is about 0.12\textperthousand.

{The distributionally robust model minimizes the expectation of scheduling cost under the worst-case distribution, and thus the objective value and the wind curtailment level in the scheduling phase are both highest. The UC decision yielded is robust against the worst-case distribution, and results in less wind curtailment in real-time operation (see the last panel in Fig.~\ref{fig:Cost_and_curt}). However, since the worst-case distribution rarely occurs, the UC solution is somewhat conservative and pessimistic. As shown in Table~\ref{tab:cost_and_curt}, the simulation cost for the distributionally robust model is highest, regardless of the lowest wind curtailment level. Another reason for the conservativeness is that the ambiguity set of DR-IEGS fails to model the correlation of random variables, and the extremal distribution contains many fast ramping events that are unlikely to occurs in reality. Although high-order moments can capture spatial and temporal correlations, incorporating them to DR-IEGS will give rise to some semidefinite programmings and bi-convex programmings, making the model more difficult to solve~\cite{zheng2019mixed}.}

\begin{table}[t]
\footnotesize
\small
\center

\caption{Comparisons of Costs and Wind Curtailments}\label{tab:cost_and_curt}
\begin{tabular}{cc|cc}
\hline
 &  & Scheduled & Expected \\ \hline
\multirow{2}{*}{D-IEGS} & Obj. (k\$) & 3,071.06 & 3,078.54 \\
 & Curt. (MWh) & 0.00 & 10.01 \\ \hline
\multirow{2}{*}{S-IEGS} & Obj. (k\$) & 3,073.26 & {\ul 3,078.19} \\
 & Curt. (MWh) & 3.04 & 7.84 \\ \hline
\multirow{2}{*}{DR-IEGS} & Obj. (k\$) & 3,085.25 & 3,079.42 \\
 & Curt. (MWh) & 36.50 & {\ul 0.75} \\ \hline
\end{tabular}
\end{table}

Throughout the computational experiment, load shedding doesn't occur in IEGS. This should be owed to the flexibility originating from gas storage stations and the line-pack effect.

\subsection{Settlement of PtGs using E-LMV}
\begin{figure}[t]
  \centering
  \includegraphics[width=3.3in]{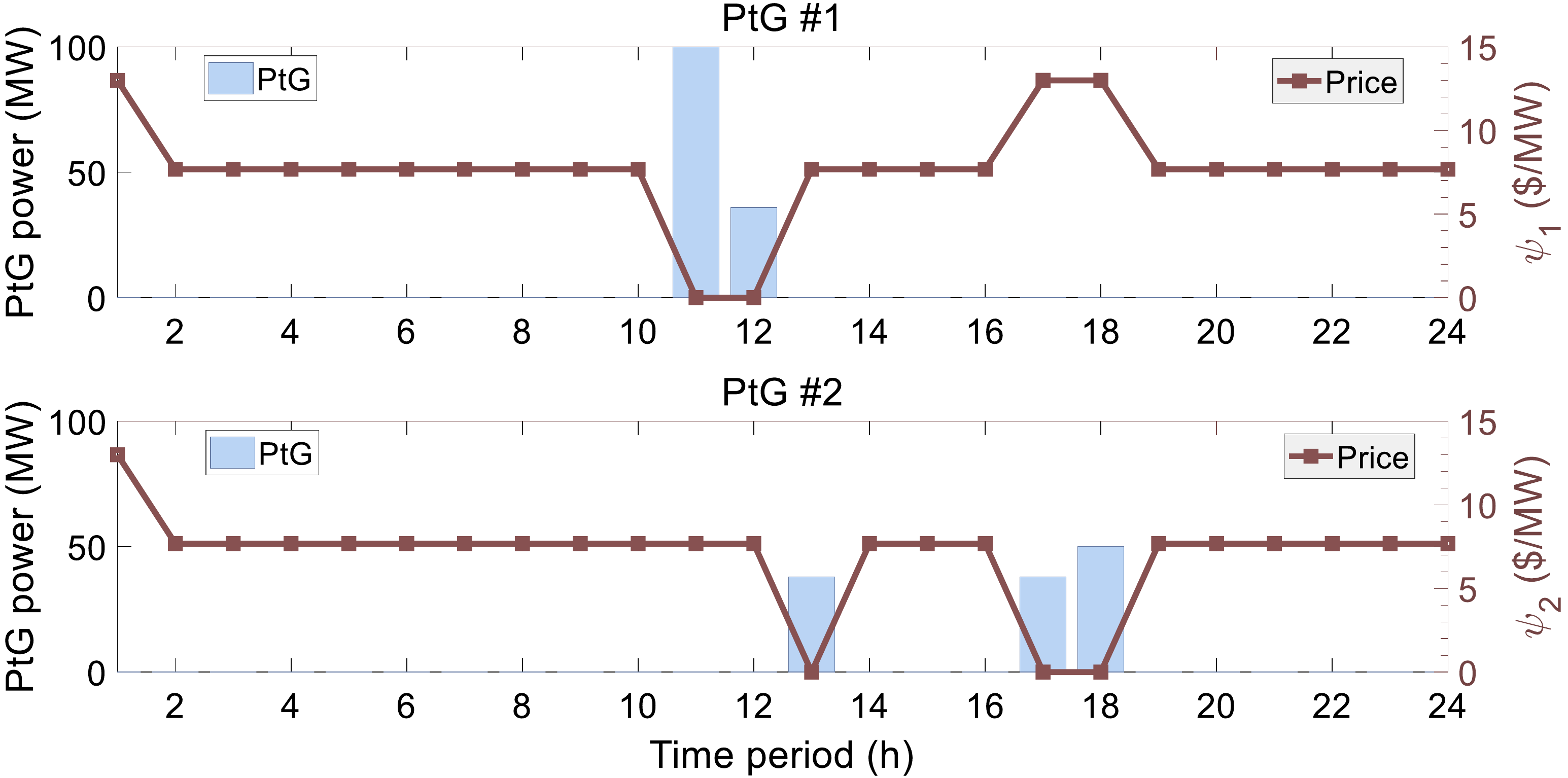}
  \caption{PtG production levels with respective to LMPs.}\label{fig:PtG_LMP}
\end{figure}

To settle the day-ahead market, as usual, the UC solution yielded from S-IEGS is fed back to the deterministic model to obtain a pre-dispatch solution and LMPs. {In this way, }the PtG production levels as optimally scheduled are presented in Fig.~\ref{fig:PtG_LMP}, together with LMPs defined by Eqn.~(\ref{eqn:LMP_PtG}).
{In the test system, LMPs of the power system range from 4.32~\$/MW to 17.32~\$/MW, while those of the natural gas system range from 7.15~\$/MW  to 7.45~\$/MW (considering the efficiency factor, it is 4.14~\$/MW to 4.32~\$/MW).}
According to Fig.~\ref{fig:PtG_LMP}, PtGs convert power to gas only when $\psi$ is \emph{zero}, that is, the LMPs on the power system side and the gas system side {all equal} 4.32~\$/MW. This verifies the claim in Section~\ref{sec:IEGS_pricing}.

{Noting that the minimum variable cost of generators is 10~\$/MW~\cite{zheng_2019},} LMPs take the value of 4.32 \$/MW only when the wind farms encounter overproduction. However, when overproduction occurs, absent PtGs, the LMPs of such buses would be non-positive.
Therefore, it is easy to see that PtGs consume excess wind power, raise up the price, and end up getting less payment and often zero payoff. In this case, the payment to PtGs derived from LMPs is 0 k\${, because congestion doesn't occur near Bus 32 and 33 under the forecast scenario}.

\begin{figure}[t]
  \centering
  \includegraphics[width=3.3in]{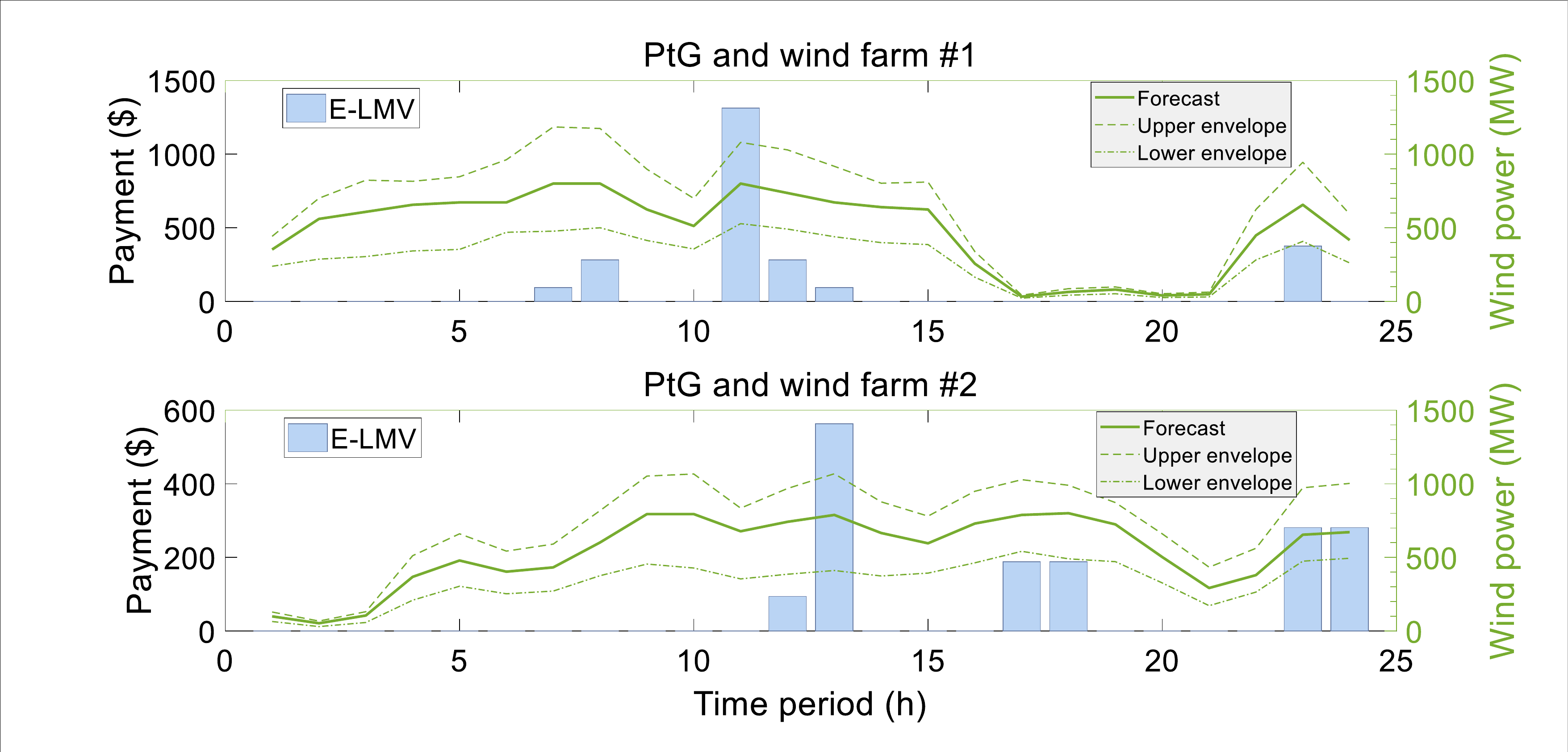}
  \caption{Payments to PtGs with respect to wind power (intervals).}\label{fig:Pay_wind}
\end{figure}

If the market is settled using E-LMVs, the payments of PtGs at each time period are as shown in Fig.~\ref{fig:Pay_wind}.
\textcolor{black}{In Fig.~\ref{fig:Pay_wind}, the day-ahead forecast and the upper/lower envelop are also plotted. The envelope is obtained by taking the pointwise maximum/minimum of wind power levels in the scenario set, so it indicates the highest/lowest possible wind power level in the stochastic model.}
{In this test system, the wind power capacity is 1200 MW, i.e., 100-MW higher than the summation of the PtG capacity and the transmission line capacity. Therefore, congestion occurs either when CfU at Bus 32 or 33 is scheduled OFF and the wind power exceeds 1100 MW, or when CfU at Bus 32 or 33 is scheduled ON and the wind power exceeds 900 MW or 950 MW (subtracting the minimum production level of the CfU).
In the stochastic model, it is hard to seek a UC solution that incurs no congestion under all probabilistic scenarios. Therefore, payments to PtGs are more likely to occur.
\textcolor{black}{It can be seen from Fig.~\ref{fig:Pay_wind} that payments occur even when the highest possible wind power level is less than 1100 MW, because congestion exists under some scenarios given the optimal UC solution. However, if the payment is derived from the forecast value (i.e., the expected scenario), the payment is zero as above-mentioned. Therefore, E-LMV better reflects the expected value of PtGs than LMP of the expected scenario does.}
It can be expected that under a same system configuration, the more volatile and uncertain wind power is, the higher E-LMV will be.}

The total credit to PtGs derived from S-IEGS is 4.03 k\$. {The value obtained from DR-IEGS is 27.82 k\$, which is several times higher than that from S-IEGS. In fact, it may not be persuasive to settle the market based on the worst-case situation.
}

As defined in Eqn.~(\ref{eqn:E-LMV_PtG}), the $\omega$-th scenario contributed to $\mathbb{E}[{\Psi}_{v,t}]$ only if $\psi_{v,t,\omega}$ is negative, which requires that $p^{\rm{PtG}}_{v,t,\omega} = \overline{p}^{\rm{PtG}}_{v,t}$.
{Hence, the mechanism of the proposed settlement scheme} is akin to the financial transmission right, but in a stochastic setting. According to Proposition~\ref{pro:1}, the payment to PtGs is balanced by the charge from volatile renewable generations and demands. The payment received by PtG owners can be spent on capacity expansion.

\subsection{Long-term Marginal Value of PtGs}
Using the same setting, we solve S-IEGS and run simulations for cases with different PtG capacities to assess the long-run contribution of PtGs.

According to Table~\ref{tab:PtG_capacity}, the marginal value of installing 100 extra MW of PtGs is remarkable when the initial capacity is 100 MW, which is given by the difference of expected costs, i.e., $(3,085.37-3,077.69)\times312/365 + (3,087.93-3,081.14)\times53/365 = 7.56$ k\$. Moreover, Table~\ref{tab:PtG_capacity} provides a straightforward alternative for evaluating the daily value of the existing 200-MW PtGs, i.e., by taking the difference of the expected costs under the 0-MW and 200-MW capacities, the daily marginal value can be obtained, which is 12.32 k\$ or about 4.00\textperthousand~of the total cost. This number has the same order of magnitude with {E-LMVs derived from S-IEGS and DR-IEGS}.
\begin{table}[t]
\footnotesize
\small
\centering
\caption{Costs of IEGS with Different PtG Capacities}\label{tab:PtG_capacity}
\begin{tabular}{c|ccc}
    \hline
    \begin{tabular}[c]{@{}c@{}}PtG capacity\\ (MW)\end{tabular} & \begin{tabular}[c]{@{}c@{}}Scheduled\\ (k\$)\end{tabular} & \begin{tabular}[c]{@{}c@{}}In-sample\\ (k\$)\end{tabular} & \begin{tabular}[c]{@{}c@{}}Out-of-sample\\ (k\$)\end{tabular} \\ \hline
    0	&	3,079.64	&	3,090.43	&	3,090.93	\\
    100	&	3,073.94	&	3,085.37	&	3,087.93	\\
    \underline{200}	&	3,073.26	&	3,077.69	&	3,081.14	\\
    300	&	3,073.26	&	3,077.18	&	3,080.37	\\
    \hline
\end{tabular}
\end{table}

Although the PtG technology is still costly, the cost saving achieved by installing such facilities can be much higher than that via optimal scheduling only (4.00\textperthousand~v.s. 0.12\textperthousand). For IEGS, it is of vital importance to decide an economic PtG size. From this perspective, the results in Table~\ref{tab:PtG_capacity} also suggest the applicability of S-IEGS model and the proposed algorithm to optimally sizing PtG capacities.

\section{Conclusions and Discussions}\label{sec:conclusion}
In this paper, a data-driven stochastic model is developed to co-optimise IEGS in day-ahead markets and address multiple correlated uncertainties.
\textcolor{black}{The data-driven stochastic model has cost benefit compared with a deterministic model. Moreover, it is demonstrated that the stochastic model has advantage over a distributionally robust model in terms of algorithmic tractability, and also on cost efficiency due to the fact that the stochastic programming framework allows more precise modeling of the gas flow problem.}

\textcolor{black}{The proposed algorithm ensures convergence and provides high-quality solutions to the original MINLP problem, even under a decentralised computational setting. The computational time is reasonable regarding the clearing time of day-ahead market, as the algorithm framework allows parallel and distributed computing.}

\textcolor{black}{According to the analysis of LMPs at coupling buses/nodes, cost recovery is difficult for PtGs under a deterministic-LMP-based regime. The expected locational marginal value proposed in this paper provides an alternative to pricing PtG facilities in a day-ahead market with production and demand uncertainties, and it ensures that PtGs get sufficient payments to expand their capacities to better mitigate the volatile renewable generations.} It is also demonstrated that the cost saving achieved by installing PtGs is higher than that via optimal scheduling.

The direction of gas flow is fixed in this model. \textcolor{black}{In future work, however, bi-directional flow will be modeled and more sophisticated algorithms should be developed~\cite{belderbos2020facilitating}.} As for the data-driven stochastic model, it is useful to improve the samples by exploiting more statistical features of historical data, or using importance sampling~\cite{papavasiliou2013multiarea},{\cite{qiu2014multi}}, etc.

\section{Acknowledgments}\label{sec:acknowledgment}
This work was supported by the National Natural Science Foundation of China (51937005).


{\color{black}
\section{Appendix}
\subsection{Proof of Proposition~\ref{pro:1}}
\begin{proof}
Since E-LMV is the weighted sum of the payments under different scenarios, we only have to prove that the payment scheme derived from LMP at each deterministic scenario ensures revenue adequacy for IEGS. Furthermore, PtG (GfU) can be regarded as the buyer (seller) in the electricity market, and the seller (buyer) in the gas market, so it is possible to fix the transactions between these two markets as the optima, and prove revenue adequacy for each individual system.

The proof is based on the Lagrangians of the optimisation models and the Karush-Kuhn-Tucker (KKT) first-order necessary conditions of optimality~\cite{gomez2008electric}. The variable after a colon represents the Lagrangian multiplier of the constraint.
\subsubsection{Revenue Adequacy of Electricity Market}
With a fixed UC decision and PtG/GfU production level, the SCUC problem~(\ref{eqn:SCUC}) becomes an LP. The dc power flow equation~(\ref{eqn:SCUC_dc_equation}) can be written in a compact matrix form:
\begin{equation} \label{eqn:scuc_dc}
  \bm{B\theta} = \bm{p}_{\rm{G}} + \bm{p}^{\ast}_{\rm{GfU}} + \bm{P}_{\rm{W}} - \bm{P}_{\rm{D}} - \bm{p}^{\ast}_{\rm{PtG}} ~ :\bm{\mu}.
\end{equation}
According to the primal feasibility condition (\ref{eqn:scuc_dc}), we have
\begin{equation}
  \bm{\mu}^{\top}\bm{B\theta} = \bm{\mu}^{\top}(\bm{p}_{\rm{G}}  + \bm{p}^{\ast}_{\rm{GfU}} + \bm{P}_{\rm{W}} - \bm{P}_{\rm{D}} - \bm{p}^{\ast}_{\rm{PtG}}).
\end{equation}
Revenue adequacy requires that the money collected from the consumers is more than that paid to the suppliers, that is, $\bm{\mu}^{\top}(\bm{P}_{\rm{D}} + \bm{p}^{\ast}_{\rm{PtG}}) - \bm{\mu}^{\top}(\bm{p}_{\rm{G}}  + \bm{p}^{\ast}_{\rm{GfU}} + \bm{P}_{\rm{W}}) \ge 0$, which in turn requires that $\bm{\mu}^{\top}\bm{B\theta} \le 0$.

The power flow constraint~(\ref{eqn:SCUC_flow_limit}) can be written as
\begin{equation} \label{eqn:scuc_flow}
\left\{
    \begin{aligned}
        \bm{X}^{-}\bm{\theta} &\le \overline{\bm{F}}  ~~ :\bm{\vartheta}^{+} \\
        -\bm{X}^{-}\bm{\theta} &\le \overline{\bm{F}}  ~~ :\bm{\vartheta}^{-}.
  \end{aligned}
\right.
\end{equation}
The complementary slackness condition of Eqn.~(\ref{eqn:scuc_flow}) is
\begin{equation} \label{eqn:kkt_scuc_flow}
\left\{
    \begin{aligned}
        \bm{\vartheta}^{+\top}(\bm{X}^{-}\bm{\theta} - \overline{\bm{F}}) &= 0 \\
        \bm{\vartheta}^{+\top}(-\bm{X}^{-}\bm{\theta} - \overline{\bm{F}}) &= 0.
  \end{aligned}
\right.
\end{equation}

The constraint for reference bus~(\ref{eqn:SCUC_ref}) is omitted without affecting the conclusion. Hence, applying the dual feasibility condition associated with the primal variable $\bm{\theta}$ leads to the following equality,
\begin{equation} \label{eqn:kkt_scuc_theta}
  (\bm{\mu}^{\top}\bm{B} + (\bm{\vartheta}^{+}-\bm{\vartheta}^{-})^{\top}\bm{X}^{-})\bm{\theta} = 0.
\end{equation}

Combining Eqn.~(\ref{eqn:kkt_scuc_flow}) and (\ref{eqn:kkt_scuc_theta}), we then have
\begin{equation}
  \bm{\mu}^{\top}\bm{B\theta} = -(\bm{\vartheta}^{+}-\bm{\vartheta}^{-})^{\top}\bm{X}^{-}\bm{\theta} = -(\bm{\vartheta}^{+}+\bm{\vartheta}^{-})^{\top}\overline{\bm{F}}.
\end{equation}
Since $\bm{\vartheta}^{+}$, $\bm{\vartheta}^{-}$ and $\overline{\bm{F}}$ are all non-negative, it concludes that $\bm{\mu}^{\top}\bm{B\theta} \le 0$, and thus the revenue adequacy of electricity market is guaranteed. The revenue of electricity market, if exists, is due to the congestion of transmission line, and hence is known as the congestion revenue.

\subsubsection{Revenue Adequacy of Gas Market}
With a fixed PtG/GfU production level, the gas balance equation~(\ref{eqn:GS_flow_balance}) becomes
\begin{equation} \label{eqn:gs_balance}
    \begin{aligned}
        \bm{f}_{\rm{Src}} + \bm{f}_{\rm{PtG}}^{\ast} = \bm{f}_{\rm{Str}} + \bm{f}_{\rm{GfU}}^{\ast} + \bm{F}_{\rm{Load}} + \bm{f}_{\rm{Cmp}} + \bm{f}_{\rm{Pipe}} ~ :\bm{\lambda}.
    \end{aligned}
\end{equation}
The primal feasibility condition of (\ref{eqn:gs_balance}) leads to
\begin{equation}
\bm{\lambda}^{\top}(\bm{f}_{\rm{Cmp}} + \bm{f}_{\rm{Pipe}}) = \bm{\lambda}^{\top}(\bm{f}_{\rm{Src}} + \bm{f}_{\rm{PtG}}^{\ast} - \bm{f}_{\rm{Str}} - \bm{f}^{\ast}_{\rm{GfU}} - \bm{F}_{\rm{Load}}).
\end{equation}
Revenue adequacy requires that $\bm{\lambda}^{\top}(\bm{f}_{\rm{Cmp}} + \bm{f}_{\rm{Pipe}}) \le 0$. However, the sign of $\bm{\lambda}^{\top}(\bm{f}_{\rm{Cmp}} + \bm{f}_{\rm{Pipe}})$ is not clear yet since $\bm{f}_{\rm{Cmp}}$ and $\bm{f}_{\rm{Pipe}}$ are free variables ($\bm{\lambda}$ might also be negative).

Hereafter, $\bm{f}_{\rm{Cmp}}$ and $\bm{f}_{\rm{Pipe}}$ are uniformly represented by $\tilde{\bm{f}}_{\rm{Pipe}}$, and the equations associated with $\tilde{\bm{f}}_{\rm{Pipe}}$, $\bm{e}$ and $\bm{\pi}$ are recast as follows,
\begin{equation} \label{eqn:kkt_gs_f_Pipe_bar_and_E}
\left\{
    \begin{aligned}
        &{\text{(\ref{eqn:GS_f_Pipe_bar_f})}} \rightarrow \bm{D}_1\tilde{\bm{f}}_{\rm{Pipe}} = \bar{\bm{f}}_{\rm{Pipe}}~ &:\bm{\varphi}_1 \\
        &{\text{(\ref{eqn:GS_cmp_lp_f}),~(\ref{eqn:GS_lp_f})}} \rightarrow \bm{D}_{2,1}\tilde{\bm{f}}_{\rm{Pipe}} = \bm{D}_{2,2}\bm{e} ~ &:\bm{\varphi}_2 \\
        &{\text{(\ref{eqn:GS_f_Pipe_bar_pi})}} \rightarrow\bar{\bm{f}}_{\rm{Pipe}} \le \bm{C}_1\bm{\pi}~ &:\bm{\zeta}_1 \\
        &{\text{(\ref{eqn:GS_lp_pi})}} \rightarrow\bm{e} = \bm{C}_2\bm{\pi}~ &:\bm{\zeta}_2 \\
        &{\text{(\ref{eqn:GS_lp_identity})}} \rightarrow\bm{C}_3\bm{e} = \bm{E}_0~ &:\bm{\zeta}_3 \\
        &{\text{(\ref{eqn:GS_pi_limit})}} \rightarrow\bm{\pi} \le \overline{\bm{\Pi}}~ &:\bm{\vartheta}_{1} \\
        &{\text{(\ref{eqn:GS_pi_limit})}} \rightarrow -\bm{\pi} \le -\underline{\bm{\Pi}}~ &:\bm{\vartheta}_{2}
    \end{aligned}
\right.,
\end{equation}
where $\bm{D}_{[\cdot]}$, $\bm{D}_{[\cdot,\cdot]}$ and $\bm{C}_{[\cdot]}$ are coefficient matrixes with appropriate dimensions. Herein, the general flow equation~(\ref{eqn:GS_f_Pipe_bar_pi}) is linearised for the sake of simplicity.

Based on the KKT conditions of Eqn.~(\ref{eqn:kkt_gs_f_Pipe_bar_and_E}), we have
\begin{equation} \label{eqn:kkt_gs}
    \begin{aligned}
    &(\bm{\varphi}_1^{\top}\bm{D}_1 + \bm{\varphi}_2^{\top}\bm{D}_{2,1}) \tilde{\bm{f}}_{\rm{Pipe}} \\
    = &\bm{\varphi}_1^{\top} \bar{\bm{f}}^{\rm{Pipe}} + \bm{\varphi}_2^{\top}\bm{D}_{2,2}\bm{e} \\
    = &\bm{\zeta}_1^{\top} \bar{\bm{f}}_{\rm{Pipe}} + \bm{\zeta}_2^{\top} \bm{e} + \bm{\zeta}_3^{\top}\bm{C}_3 \bm{e} \\
    =  &\bm{\zeta}_1^{\top} \bm{C}_1\bm{\pi} + \bm{\zeta}_2^{\top} \bm{C}_2\bm{\pi} + \bm{\zeta}_3^{\top} \bm{E}_0 \\
    = &\bm{\vartheta}_{1}^{\top}\overline{\bm{\Pi}} - \bm{\vartheta}_{2}^{\top}\underline{\bm{\Pi}} + \bm{\zeta}_3^{\top} \bm{E}_0 \ge 0.
    \end{aligned}
\end{equation}
For a general gas system, it's reasonable to assume that $\underline{\bm{\Pi}}$ and $\bm{E}_0$ are both $\bm{0}$ (otherwise specific operation data is needed to analyze the revenue), and hence the inequality in Eqn.~(\ref{eqn:kkt_gs}) holds given that $\bm{\vartheta}_{1}$ and $\overline{\bm{\Pi}}$ are both non-negative.

Moreover, the dual feasibility condition associated with $\tilde{\bm{f}}_{\rm{Pipe}}$ suggests that
\begin{equation}\label{eqn:kkt_gs_f_Pipe}
  (\bm{\lambda}^{\top} + \bm{\varphi}_1^{\top}\bm{D}_1 + \bm{\varphi}_2^{\top}\bm{D}_{2,1}) \tilde{\bm{f}}_{\rm{Pipe}} = 0,
\end{equation}
where $\bm{\lambda}^{\top} \tilde{\bm{f}}_{\rm{Pipe}} = \bm{\lambda}^{\top}(\bm{f}_{\rm{Cmp}} + \bm{f}_{\rm{Pipe}})$ by construction.

The conclusion that $\bm{\lambda}^{\top} \bm{f}_{\rm{Pipe}} \le 0$ now can be drawn based on Eqn.~(\ref{eqn:kkt_gs}) and (\ref{eqn:kkt_gs_f_Pipe}). The revenue, if exists, is caused by the limitations of flow rate and line-pack capacity, which are determined by $\overline{\bm{\Pi}}$,
\end{proof}
\subsection{Discussions}
In fact, it is due to the enforced nodal pressures/flow rates, instead of the gas loss, that the revenue adequacy of gas market cannot be verified when compressors exist. This is similar to the electricity market. For example, if the rate of power flow on a transmission line is enforced to be higher than some levels, then costly power may flows to less-expensive locations, and the revenue adequacy of electricity market is not guaranteed.

For a nonlinear gas market with gas compressors, the revenue might still be non-negative in reality though it cannot be verified in theory. This is in line with the observation in numerical experiments, i.e., with the optimal Lagrangian multipliers of the SOCP model, one of the source nodes (Node 8) has a lower gas price than those at demand nodes, guaranteeing that the cost of the gas consumed by compressors can be compensated precisely (the revenue adequacy of the gas market is zero).
}

\end{document}